\documentclass[letterpaper]{article}
\usepackage{spconf}
\usepackage[T1]{fontenc}
\usepackage{amsmath,amssymb}
\usepackage{graphicx}
\usepackage{pgfplots}
\pgfplotsset{compat=1.17}
\usepgfplotslibrary{groupplots}
\usepackage{booktabs}
\usepackage{tabularx}
\usepackage{xcolor}
\usepackage[final]{microtype}
\usepackage[hidelinks]{hyperref}
\hypersetup{
  pdftitle={Robust, Estimator-Agnostic Dynamic 3DGS Compression},
  pdfauthor={Chenjunjie Wang, Zixi Huang, Yao Wang, and Jona Ballé},
  pdfkeywords={3D Gaussian splatting, dynamic scene compression,
    post-training compression, Gaussian splat coding}
}
\usepackage{cite}
\usepackage{enumitem}
\usepackage{etoolbox}

\ifdefined\NOAPPENDIX
  \apptocmd{\thebibliography}{%
    \fontsize{9}{9}\selectfont
    \setlength{\itemsep}{0pt}
    \setlength{\parsep}{0pt}
  }{}{}
\else
  \apptocmd{\thebibliography}{%
    \fontsize{9}{8.6}\selectfont
    \setlength{\itemsep}{-0.25ex}
    \setlength{\parsep}{0pt}
  }{}{}
\fi

\ifdefined\NOAPPENDIX
  \newcommand{\appx}[1]{}
  
\else
  \newcommand{\appx}[1]{#1}
  
\fi
\newcommand{\solidkey}{\raisebox{0.28ex}{\rule{1.35em}{0.9pt}}}
\newcommand{\dashkey}{\raisebox{0.28ex}{\rule{0.33em}{0.9pt}\hspace{0.17em}%
  \rule{0.33em}{0.9pt}\hspace{0.17em}\rule{0.33em}{0.9pt}}}
\newcolumntype{Y}{>{\raggedright\arraybackslash}X}

\allowdisplaybreaks[2]
\begin{document}
% Official spconf option: 9 pt text, within the ICASSP minimum.
\ninept

\title{Robust, Estimator-Agnostic Dynamic 3DGS Compression}

\name{
Chenjunjie Wang \qquad
Zixi Huang \qquad
Yao Wang \qquad
Jona Ball\'e
\appx{\thanks{
This work has been submitted to the IEEE for possible publication.
Copyright may be transferred without notice.
}}
}

\address{
New York University\\
\{cw4287, zh2996, yw523, jona.balle\}@nyu.edu
}

\maketitle

\begin{abstract}
Dynamic 3D Gaussian splats (3DGS) model time-varying scenes using a separate Gaussian set per frame. While neighboring video frames are highly correlated due to smooth motion, Gaussian representations retain this correlation to varying degrees, depending on whether the estimator tracks them across time. Some 3DGS compression methods integrate the estimation to exploit temporal redundancy; here, we focus on robust compression regardless of the estimator. We concatenate groups of frames into one Gaussian set, augment each Gaussian with a frame index, and pass it to a static (i.e., non-temporal) 3DGS codec, converting temporal redundancy into spatial redundancy. Concatenated sets are spatially partitioned to limit memory. Our technique requires neither a motion model nor knowledge of the training method. Averaged over six N3DV sequences, all six static codecs achieve gains on tracked sets ($-42.0\%$ to $-71.8\%$ BD-rate) over per-frame coding. On untracked sets, all codecs except HGSC, which appears incompatible with our technique, remain competitive with per-frame coding ($-3.5\%$ to $+5.0\%$). We further replace D-FCGS’s I-frame coding with our technique while retaining its P-frame coding, yielding an overall BD-rate of $-46.2\%$. We propose to visualize “trackedness” using an inter-frame similarity metric. The project is available at \url{https://wcjj1236.github.io/d3dgs-benchmark}.
\end{abstract}


% ---------------------------------------------------------------------
\section{Introduction}
3D Gaussian Splatting (3DGS) has recently emerged as a significant 3D scene representation (historically, ``scene reconstruction''), combining photorealistic visual quality with real-time rendering~\cite{kerbl3dgs}. For  each scene, a set of Gaussian primitives located in 3D space and with view-dependent rendering properties is estimated from multiple training views, enabling novel-view synthesis.  Different methods exist for adding a time dimension to 3DGS: simply estimating a
new set of Gaussians for every temporal frame \cite{kerbl3dgs,wei2026dancenet3d} (``untracked'' estimation), estimating a `canonical' set of Gaussians initially and deforming it over time (``tracked'' estimation)~\cite{wu4dgaussians,yang4dgs}, propagating and selectively updating a set of Gaussians (``semi-tracked'' estimation)~\cite{sun3dgstream,girish2024queen} or emitting a frame-wise 3D Gaussian sequence using feed-forward networks~\cite{ren2024l4gm}. Despite these different estimation processes, each method eventually produces a set of 3D Gaussian primitives for each frame, albeit with different temporal correlation patterns. We consider the problem of compressing these
frame-wise 3D Gaussian sequences independently of the estimation method, and with an unknown correlation pattern.

\begin{figure*}[t]
\centering
\includegraphics[width=0.96\textwidth]{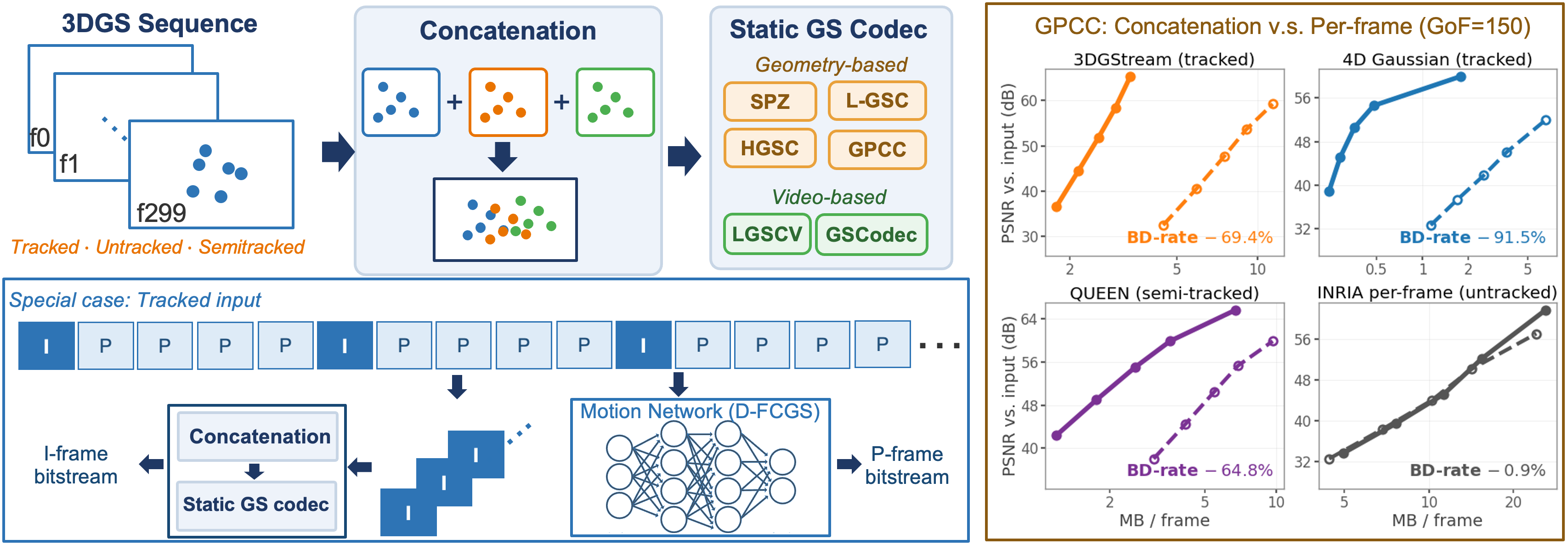}
\caption{Overview of the proposed method. Left: GoF
concatenation is passed to a geometry- or video-based static GS codec;
on tracked input, the figure illustrates the concatenation branch of our D-GPCC which replaces the D-FCGS's intra branch while retaining
P-frame motion coding. Right: N3DV-average G-PCC rate--distortion curves
at GoF-150, comparing \solidkey~concatenation with
\dashkey~per-frame coding branches.}
\label{fig:overview}
\end{figure*}

Our approach is simple: We concatenate each group of frames (GoF) into a single Gaussian set and compress it with a static 3DGS codec (Fig.~\ref{fig:overview}), optionally slicing it spatially to control memory usage. Each Gaussian's frame is recovered exactly (Sec.~\ref{sec:experiments}). Our method allows the static codec to exploit temporal correlations within the concatenated set, the same way it exploits spatial correlations. The method requires no cross-frame Gaussian correspondence, reference frames, or motion fields, regardless of the origin of the Gaussian sets.

To assess the utility of our idea across architectures, we evaluate it using six different static GS codecs and ask whether the proposed conversion achieves gains in each setting. For each codec, concatenated coding is compared with a per-frame baseline using the same codec.
We find that in each case except one, our method is at least competitive with the baseline, and can improve compression efficiency substantially, especially for tracked and semi-tracked sequences. We also integrate our method into D-FCGS by replacing its FCGS-based I-frame coding with concatenated G-PCC while retaining its original P-frame coding, yielding a $46.2\%$ overall BD-rate reduction. Finally, we propose an inter-frame similarity metric which captures the correlation pattern of each GoF and may help to predict which codecs and sequences will benefit most from concatenation.

% Declared early so the double-column float lands at the top of page 3.

\begin{figure}[!b]
\centering
\includegraphics[width=\columnwidth]{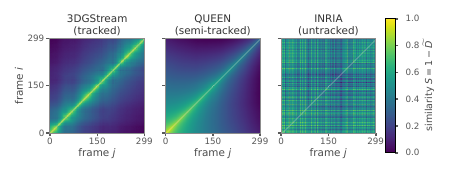}
\caption{Pairwise inter-frame similarity matrices for \emph{Flame Steak},
one example per correlation pattern we observed depending on the estimation method. Brighter entries denote greater similarity: tracked and semi-tracked sequences retain clear temporal structure, whereas the untracked sequence does not.}
\label{fig:wd}
\end{figure}

\section{Proposed Method}
\label{sec:background}
\label{sec:method}

% \subsection{Sequence Types and Evaluation}
% \label{sec:types}
% \label{sec:sequences}
A frame-wise sequence of Gaussians is $\mathcal{G}=\{G_t\}_{t=1}^{T}$, where
\begin{equation}
G_t=\left\{g_{t,n}=
(\boldsymbol{\mu}_{t,n},\mathbf{\Sigma}_{t,n},\mathbf{SH}_{t,n},o_{t,n})
\right\}_{n=1}^{N_t}.
\label{eq:gaussian-frame}
\end{equation}
Here $\boldsymbol{\mu}$ is the Gaussian center, $\mathbf{\Sigma}$ is the covariance,
$\mathbf{SH}$ is the spherical-harmonic coefficients, and $o$ is the opacity.
Depending on the estimation method, we distinguish three types of temporal correlation patterns:
\begin{itemize}
\item tracked estimation: $N_t$ is constant and the index $n$ identifies the same
Gaussian at every $t$;
\item semi-tracked estimation: only some Gaussians persist across
frames, but $N_t$ varies over time; and
\item untracked estimation: each frame is independently reconstructed as an
unordered set, so no correspondence can be assumed across frames; only scene-level geometry and attribute
statistics are shared.
\end{itemize}

\subsection{Correlation Pattern Analysis}
\label{sec:correlation_pattern}
We visualize the different types of correlation patterns using an inter-frame similarity matrix (Fig.~\ref{fig:wd}). To construct the matrix, we first compute the pairwise squared 2-Wasserstein distance between Gaussians~\cite{givens1984wasserstein}:
\begin{equation}
\begin{aligned}
W_2^2(g_i,g_j)
&=\lVert\boldsymbol{\mu}_i-\boldsymbol{\mu}_j\rVert_2^2\\
&\quad+\operatorname{tr}\!\left(
\mathbf{\Sigma}_i+\mathbf{\Sigma}_j-
2(\mathbf{\Sigma}_i^{1/2}\mathbf{\Sigma}_j
\mathbf{\Sigma}_i^{1/2})^{1/2}
\right).
\end{aligned}
\label{eq:w2}
\end{equation}
We combine this with differences in SH coefficients and opacity to define a pairwise cost:
\begin{equation}
\begin{aligned}
C(g_i,g_j)
&=\alpha W_2^2(g_i,g_j)\\
&\quad+\beta\operatorname{MSE}(\mathbf{SH}_i,\mathbf{SH}_j)
+\gamma\operatorname{MSE}(o_i,o_j).
\end{aligned}
\label{eq:wdcost}
\end{equation}
For each Gaussian $g_{t,i}\in G_t$, let $\mathcal{N}_3^u(i)$ denote
its three nearest Gaussians in $G_u$, determined by Euclidean distance
between their centers. The frame-to-frame cost is
\begin{equation}
D(t,u)=\frac{1}{3N_t}
\sum_{i=1}^{N_t}\sum_{j\in\mathcal{N}_3^u(i)}
C(g_{t,i},g_{u,j}).
\label{eq:framecost}
\end{equation}
We define the inter-frame similarity as
$S(t,u)=1-\widetilde D(t,u)$, where $\widetilde D$ is the normalized
version of $D$ (Fig.~\ref{fig:wd}).
In practice we evaluate $C$ and $D$ on representative subsets of Gaussians (level-of-detail neighbourhoods weighting each residual by its quantization step \appx{(App.~\ref{app:similarity})}).

\subsection{Coding by Concatenation}
\label{sec:concat}\label{sec:impl}
We code each group of frames (GoFs) as one unit. For example, we indicate partitioning a $300$-frame sequence into two groups as ``GoF-150''. For a GoF $\mathcal{T}_k$ starting at $t_k$, we concatenate all Gaussians without merging duplicates, attach the relative frame index $f_i=t-t_k$ to each Gaussian from frame $t$, and form
\begin{equation}
G_k^{\mathrm{cat}}
=\biguplus_{t\in\mathcal{T}_k}G_t,
\qquad
f_i=t-t_k \quad \text{for } g_i\in G_t.
\label{eq:concat}
\end{equation}
The GS codec receives one concatenated static Gaussian set. Concatenation increases the Gaussian count, but also allows the static codec to exploit correlation in geometry and attributes across frames. The frame indices must be transmitted losslessly to recover each output frame.
If a GoF exceeds a codec's point or memory limit, we spatially partition the concatenated set into independently coded slices, which does not require modifying Eq.~\eqref{eq:concat}.

\begin{figure*}[t]
\centering
\includegraphics[width=\linewidth]{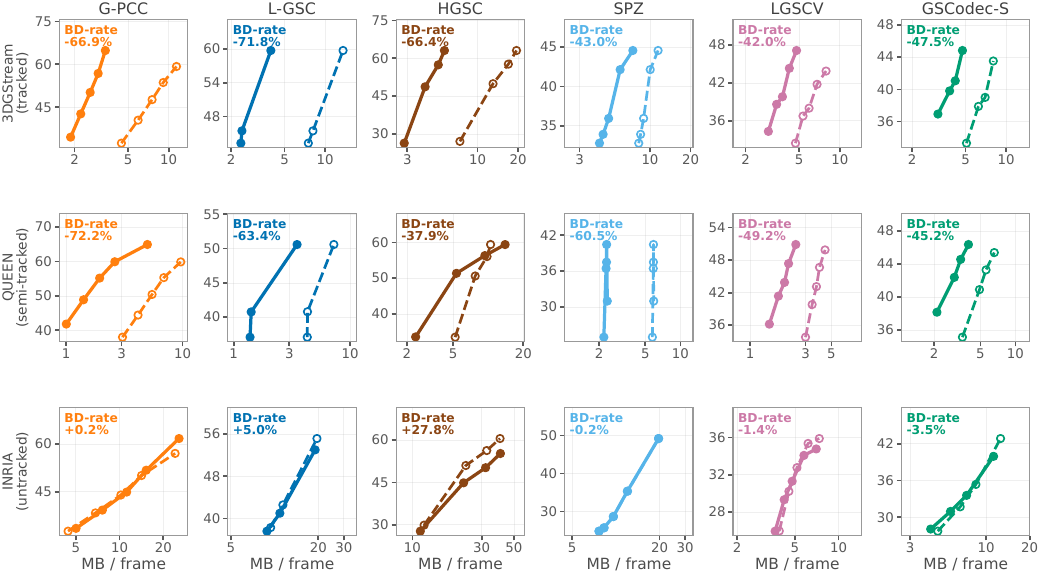}
\caption{Rate vs. distortion compression results with six codecs at GoF-30, averaged over
the N3DV dataset. Distortion is measured in PSNR with respect to the views rendered from the estimated Gaussians before compression. The three rows show examples of tracked, semi-tracked, and untracked estimators. Each panel compares \solidkey~concatenation with \dashkey~per-frame coding and reports
BD-rate.}
\label{fig:avg-grid}
\end{figure*}

We consider a number of different underlying static GS codecs to evaluate the usefulness of our concatenation method in the context of different coding systems. G-PCC \cite{mpegtmc13,graziosi2020overview,liu2020mpegpcc}
exploits spatial redundancy through octree geometry coding
\cite{huang2006octree} and attribute coding based on RAHT
\cite{dequeiroz2016raht}, predictive, or lifting with adaptive
quantization \cite{bhatt2019lifting,wang2021adaptivequant}.
HGSC \cite{huang2024hgsc} combines octree geometry coding with
hierarchical attribute prediction. L-GSC \cite{lgsc} applies
lossless compression to Morton-ordered quantized data, while
SPZ \cite{spz} compresses quantized attribute byte streams.
Concatenation is performed before codec-specific quantization. Similar
Gaussians may then share quantized spatial support, reduce prediction
residuals, or produce repeated quantized symbols for more efficient entropy coding.

For video-based GS codecs, we jointly map the concatenated Gaussians
to spatially smooth 2D attribute maps. LGSCV \cite{lgscv} sorts the
concatenated Gaussian set by 3D Morton code and refines the layout of 2D attribute maps with MiniPLAS,
while GSCodec-S \cite{liu2025gscodec} applies PLAS
\cite{morgenstern2024sog} sorting directly to the 2D attribute maps of the concatenated set. In our implementation,
both encode the sorted attribute maps using HEVC intra coding
\cite{sullivan2012hevc}. Joint sorting can place Gaussians with similar
geometry and attributes from different frames at nearby image locations.
SPZ keeps the Gaussian order and needs no frame index; L-GSC, LGSCV and
GSCodec-S reorder the Gaussians and are charged for it (Sec.~\ref{sec:exp5}).
This allows the static codecs to exploit inter-frame redundancy even in untracked sequences.

\subsection{Concatenated Key Frame Coding}
\label{sec:integration}
D-FCGS~\cite{zhang2025dfcgs} uses an I-P coding structure, in which an intra-coded key frame opens each GoF, and each later frame is predicted from the one before it using a motion network. Here, we replace the separate intra codec used for the I-frames with our concatenation method. We encode the concatenated key frames with G-PCC, while leaving the motion network and all P-frame operations unchanged. We refer to the resulting hybrid codec as D-GPCC. In spite of their temporal separation, the key frames are still strongly correlated because they come from the same sequence, and their amortized bit rate dominates the compressed size; reducing only this intra component can therefore improve the codec substantially overall.

\section{Experiments}
\label{sec:experiments}

All experiments use the six 300-frame N3DV sequences
\cite{li2022n3dv} at $1352\!\times\!1014$, with $18$--$21$ calibrated
views per sequence. Estimators include tracked 4DGaussians\cite{wu4dgaussians}, the tracked
base stream of 3DGStream \cite{sun3dgstream}, semi-tracked QUEEN\cite{girish2024queen}, and untracked per-frame INRIA 3DGS\cite{kerbl3dgs,wei2026dancenet3d}. We use the 3DGStream base stream as our primary tracked input because including spawned Gaussians would make
the sequence semi-tracked and exclude codecs requiring full
correspondence; its frame-wise export introduces less distortion than 4DGaussians. For the untracked input, each INRIA
3DGS frame is estimated independently from the same calibrated camera
poses.

For each static codec, we compare concatenation with a per-frame coded baseline using identical input sequences, rate settings, and views. BD-rate follows~\cite{bjontegaard2001}. Rates include the frame
index (Sec.~\ref{sec:exp5}). We measure compression
distortion by rendering the input and decoded Gaussian sets from the
same views and comparing the resulting images. Using held-out
ground-truth views instead would also include estimation and
view-generalization errors, obscuring differences caused solely by
compression. Concatenated sets exceeding a codec's point budget are split along their longest spatial axis into slices of roughly equal population, capped at about $1.1$ million Gaussians for G-PCC, without changing the GoF.

For concatenated key frame coding, we only replace the intra codec; the P-frame codec and its parameters are left unchanged. GoF length and codec-specific rate controls are ablated once, then fixed across all reported sequences.

Fig.~\ref{fig:avg-grid} illustrates the performance of concatenated coding using the same protocol over different codecs and estimators. We find that generally, performance gains are consistent with our correlation pattern analysis in Sec.~\ref{sec:correlation_pattern}.

\begin{figure}
\centering
\includegraphics[width=\linewidth]{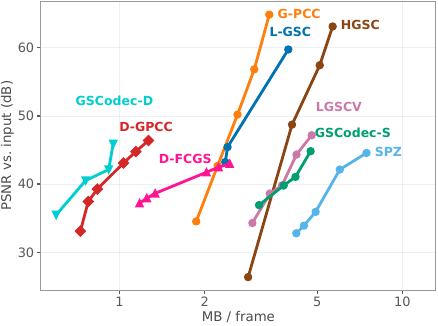}
\caption{Rate--distortion results averaged over N3DV dataset with 3DGStream (tracked)
input using concatenation. GSCodec-D, D-GPCC, and D-FCGS
require tracked input and are inherently lossy.}
\label{fig:concat-all}
\end{figure}

\subsection{Tracked and Semi-Tracked Sequences}
\label{sec:exp1}
\label{sec:exp4}
The top row of Fig.~\ref{fig:avg-grid} compares concatenated and
per-frame coding on separate axes for each codec for tracked estimation. All six concatenated
curves shift toward lower rates over overlapping PSNR ranges, with
BD-rate gains from $-42.0\%$ (LGSCV) to $-71.8\%$ (L-GSC).
Fig.~\ref{fig:concat-all} compares concatenated results on a shared axis
without per-frame baselines. It also includes three methods only applicable to
tracked input: D-FCGS~\cite{zhang2025dfcgs}, which predicts each P-frame from its predecessor
using fixed Gaussian indices; GSCodec-D~\cite{liu2025gscodec}, which reuses the I-frame's
Gaussian ordering for inter-frame coding of attribute maps. Our
D-GPCC replaces D-FCGS's I-frame coding with concatenated G-PCC.
These three methods require
full Gaussian correspondence and exploit it efficiently at low rates,
but cannot accommodate Gaussian additions and removals (semi-tracked) or independently
estimated frames (untracked); their maximum PSNR is limited relative to concatenated
G-PCC and L-GSC under the tested lossy configurations.
Our concatenated key frame coding method D-GPCC achieves a $46.2\%$ overall BD-rate reduction over D-FCGS, albeit not reaching the performance of GSCodec-D.

The middle row of Fig.~\ref{fig:avg-grid} shows similar gains for
semi-tracked sequences, ranging from $-37.9\%$ (HGSC) to $-72.2\%$
(G-PCC). The codec rankings differ: L-GSC leads on tracked
input, whereas G-PCC leads on semi-tracked input, showing that the
benefit of concatenation depends on both the codec and estimator. This
difference is also reflected in the spatial overlap across frames. At
$\mathrm{qp}{=}18$, $33.7\%$ of the Gaussians in a concatenated
\emph{Flame Steak} GoF-$30$ share a quantized cell with another Gaussian
for 3DGStream, compared with $89.2\%$ for QUEEN. Greater overlap exposes
more redundancy for a static codec to exploit.

\vspace{-0.5em}

\subsection{Untracked Sequences}
\label{sec:exp5}
For independently estimated frames, only $1.0\%$ of Gaussians share a
quantized cell, so the bottom row of Fig.~\ref{fig:avg-grid} shows smaller
and more codec-dependent changes.
Concatenated G-PCC and SPZ nearly
coincide with their per-frame baselines ($+0.2\%$ and $-0.2\%$), LGSCV
and GSCodec-S gain slightly ($-1.4\%$ and $-3.5\%$), while L-GSC and HGSC
move to higher rate ($+5.0\%$ and $+27.8\%$).
HGSC is the only evaluated codec that concatenation makes substantially worse. Its quantizers are scaled to the data they receive, so the wider spread of an untracked concatenation turns into distortion rather than rate \appx{(App.~\ref{app:untracked})}.

G-PCC and HGSC code the frame index losslessly as one $8$-bit attribute
with the predictive transform. Prediction does not help, as neighbours
mostly belong to other frames: G-PCC spends $7.7$--$8.3$ and HGSC
$9.0$--$9.9$ bits per Gaussian. Its fraction as a total of the bit rate varies: on \emph{Flame Steak} it is $2$--$8\%$ of the concatenated G-PCC
bitstream with third-order SH, against $12$--$23\%$ with first-order SH \appx{(App.~\ref{app:frameindex})}. Excluding it changes the BD-rate of concatenated G-PCC on untracked \emph{Flame Steak} from $-0.4\%$ to $-4.7\%$.
For L-GSC, LGSCV and GSCodec-S we don't actually encode the frame index, but instead add $\log_2 30=4.91$ bits per Gaussian to the plotted rates, which corresponds to $1$--$15\%$ of their total rate \appx{(App.~\ref{app:note-recovery})}.
SPZ does not require us to encode the frame index at all, since the encoding is order-preserving.

\subsection{GoF Length, Throughput, and Delay}
\label{sec:ablation}
Table~\ref{tab:gof-full} varies the GoF length for G-PCC on the tracked
3DGStream \emph{Flame Steak} sequence, where GoF-$1$ is per-frame coding.
Most of the gain comes from grouping just a few frames: GoF-$5$ already
reaches $-47.7\%$ BD-rate and GoF-$30$ $-63.4\%$, while GoF-$150$ adds
less than two points. Grouping improves throughput somewhat (Enc\,T, Dec\,T) but delays the first frame by $L/30$\,s (Delay). The concatenation algorithm needs to hold the Gaussian coordinates for the entire GoF in memory (\emph{Concat}), which grows $52\times$ by GoF-$150$. The actual encoding requires all Gaussian attributes in memory for one slice at a time (\emph{Encode}) and settles near $2.9\times$. GoF-$30$ keeps most of the
gain at a quarter of that memory.

\begin{table}[htbp]
\centering
\caption{Full GoF-length sweep, 3DGStream Flame Steak. \emph{Concat} is the
pass that holds the whole GoF, \emph{Encode} the coder on one slice.}
\label{tab:gof-full}
\footnotesize
\setlength{\tabcolsep}{2.5pt}
\begin{tabular}{ccccccc}
\toprule
GoF & \shortstack{BD-rate\,\%\\$\downarrow$} &
\shortstack{Enc\,T\\(s/f)$\downarrow$} & \shortstack{Dec\,T\\(s/f)$\downarrow$} &
\shortstack{Concat\\(GB)$\downarrow$} & \shortstack{Encode\\(GB)$\downarrow$} &
\shortstack{Delay\\(s)$\downarrow$}\\
\midrule
$1$ & Baseline & 11.7 & 9.0 & $0.06$ & $1.29$ & $0.03$\\
$5$ & $-47.72$ & 9.3 & 7.3 & $0.16$ & $3.19$ & $0.17$\\
$10$ & $-55.87$ & 8.9 & 7.0 & $0.30$ & $3.43$ & $0.33$\\
$30$ & $-63.42$ & 8.6 & 6.9 & $0.80$ & $3.49$ & $1.0$\\
$75$ & $-64.59$ & 8.7 & 6.9 & $1.73$ & $3.70$ & $2.5$\\
$150$ & $-65.05$ & 8.6 & 6.8 & $3.38$ & $3.75$ & $5.0$\\
\bottomrule
\end{tabular}
\end{table}

\section{Related Work and Discussion}
Several dynamic 3DGS compression methods jointly estimate highly compressible, time-varying Gaussian representations directly from multi-view video, e.g. QUEEN, STG, 4DGC, Light4GS, GIFStream, E-D3DGS, and
HiCoM~\cite{girish2024queen,li2024stg,
hu4dgc,liu2025light4gs,li2025gifstream,bae2024ed3dgs,gao2024hicom}.
Some static 3DGS compressors also integrate estimation~\cite{lee2024compact,chen2024hac,fan2024lightgaussian}. Post-estimation codecs based on pruning, quantization, spatial prediction, transforms, and entropy coding include \cite{huang2024hgsc,
chen2025fcgs,tian2025flexgaussian,wang2025adaptivevox,
sridhara2025ralhe}; see
\cite{bagdasarian2025survey,
mpeg2025gsc} for recent overviews.

Joint estimation--compression methods for dynamic 3DGS define the state of the art, but require re-estimating Gaussian representations from multi-view video, which may not be feasible in all circumstances.
Here we are interested in post-estimation compression, which is why we use reconstructed views from the estimated 3DGS representation as a reference rather than the ground-truth unseen views. This allows a more direct assessment of the compression method, and avoids substantial noise in the evaluation.

Our method is surprisingly simple yet effective. It shows substantial improvements on dynamic 3DGS sequences stemming from tracked and semi-tracked estimators. It is robust in the sense that it remains competitive on untracked sequences with all but one of the evaluated underlying static codecs. In addition, we find that our technique can be used on intra-frame coding with substantial benefits.

Our method builds on the underlying static GS codec's quantization, prediction, ordering, and entropy coding. Therefore, improvements to these components can also further benefit dynamic 3DGS sequence compression via concatenation. In particular, preliminary results suggest that using the GPCC-GS tools currently under development in MPEG as the intra coder of D-GPCC would substantially outperform GSCodec-D.

\section{ACKNOWLEDGMENT}

The initial concepts of concatenation and inter-frame similarity were
developed during Chenjunjie Wang's internship at Sony in collaboration with
Alexandre Zaghetto and Danillo Bracco Graziosi. The subsequent extensions,
experiments, and follow-up study were conducted at NYU after the internship.

% ---------------------------------------------------------------------
% References may begin on page 4; page 5 contains references only.
\bibliographystyle{IEEEbib}
\bibliography{refs}

% Two builds share this source: `pdflatex main` gives main text + appendix,
% and main_noappendix.tex defines \NOAPPENDIX first to stop here, so the
% submission-length version never drifts from the full one.
\ifdefined\NOAPPENDIX
  \end{document}
\fi

\clearpage

\appendix

\begin{center}
{\Large\bfseries APPENDIX}
\end{center}
\vspace{0.5em}

\section{Scope of the Appendix}
\label{app:scope}
The paper comes with two further resources, and each serves a different
purpose (Table~\ref{tab:app-resources}). This appendix defines how the
numbers in the main text are obtained, qualifies the statements of the
main text that need it (App.~\ref{app:notes}), and gives the analyses and
summary tables behind its claims. It describes each procedure at the level
needed to interpret a result, but contains no per-sequence curves and no
implementation-level change lists. The code repository, linked from the
project page, contains what is needed to rerun the experiments, and the
project page shows every measured curve interactively. The appendix, the
repository and the project page read their numbers from the same result
tables.

\begin{table}[!ht]
\centering
\caption{The three resources and what each contains.}
\label{tab:app-resources}
\footnotesize
\setlength{\tabcolsep}{3pt}
\begin{tabularx}{\columnwidth}{@{}lY@{}}
\toprule
Resource & Contents\\
\midrule
Appendix & notes on statements of the main text; sequences and estimators;
evaluation protocol; how each static GS codec handles a concatenated set;
implementation of the inter-frame similarity; mechanisms behind the gains;
frame-index overhead; construction of D-GPCC; per-sequence BD-rates; an
end-to-end comparison with published results\\
\addlinespace
Code repository & estimator training and export recipes; pinned upstream
version and every modification of each codec; codec configurations and
rate points; coding drivers; scoring code; the coding time and peak memory
of every run; all result tables with their provenance, and a script that
recomputes the numbers of the paper\\
\addlinespace
Project page & every rate--distortion curve for each sequence, codec,
estimator and GoF length; PSNR in RGB, YUV and per plane, SSIM, and the
ground-truth reference; the BD-rate of each curve against its own per-frame
branch; inter-frame similarity matrices of all four estimators\\
\bottomrule
\end{tabularx}
\end{table}

% ---------------------------------------------------------------------
\section{Notes on the Main Text}
\label{app:notes}
This section discusses statements of the main text that need
qualification, states what holds, and quantifies the effect on the
reported results. The later sections give the underlying details.

\subsection{Frame indices in the rates of L-GSC, LGSCV and GSCodec-S}
\label{app:note-recovery}
As stated in Sec.~\ref{sec:experiments}, G-PCC and HGSC code the frame
index losslessly in their bitstreams (App.~\ref{app:frameindex}), and SPZ
only needs the number of Gaussians per frame, so the rates of these codecs
include everything required to recover the frames. L-GSC, LGSCV and
GSCodec-S reorder the Gaussians and have no channel for a per-Gaussian
index, so we recover the frame of each decoded Gaussian from the reordering
computed at the encoder (App.~\ref{app:recovery}). This information is not
transmitted; in its place, the concatenated branches of the three codecs are
charged what transmitting it would cost.

The charge is the rate of a separate index stream. The decoder of each of
the three codecs outputs the Gaussians in a fixed order that it knows:
Morton order for L-GSC and the pixel order of the 2D maps for LGSCV and
GSCodec-S. An encoder can therefore write the frame index of every Gaussian
in that order and code the sequence with an arithmetic coder and a uniform
model over the $30$ frames of a GoF. Every symbol then costs exactly
$\log_2 30=4.91$ bits, whatever the data, so the rate of this stream is
known without running it: $4.91$ bits per Gaussian, plus a few bytes to
terminate the code (a fixed-length code would need $5$ bits). We add this
rate to every point of the concatenated branch; it is the mean number of
Gaussians per frame of each sequence times $4.91$ bits, between $0.19$ and
$0.29$\,MB per frame on average over the six sequences of each estimator.
This is $1.4$--$3.2\%$ of the rate of L-GSC on untracked input, $3.6$--$7.6\%$
of LGSCV and $2.3$--$6.6\%$ of GSCodec-S, and between $4.6$ and $14.6\%$ on
tracked and semi-tracked input, where the rest of the bitstream is smaller.
The per-frame branches need no index and are unchanged.

The index is hard to code for reasons that do not depend on the codec. It
must be lossless, since a wrong value assigns a Gaussian to the wrong frame,
which rules out the lossy planes of a video codec. It has no spatial
coherence by construction: concatenation pays off because co-located
Gaussians of different frames become neighbours, and the index is exactly
what tells them apart, so predicting it from neighbours does not help;
G-PCC and HGSC, which do so, spend $7.7$--$9.9$ bits per Gaussian
(App.~\ref{app:frameindex}). Its cost is the same at every rate point, so
it weighs most at low rates and on inputs whose other attributes compress
well. Only two designs avoid it: keeping the order of the Gaussians, as SPZ
does, or keeping the time axis, as GSCodec-D does on tracked input. A code
below $4.91$ bits would have to use context that the decoder already has,
such as the ordering of the codec: the Morton sort of L-GSC is stable, so
co-located Gaussians of successive frames follow each other in frame order.
We did not evaluate such a code and leave its design open.

Table~\ref{tab:app-note-recovery} shows how much the charge matters.
Without it, the BD-rates measure only what concatenation changes in each
codec's own coding of geometry and attributes; at $8$ bits per Gaussian,
close to what G-PCC spends, they approach the cost of coding the index as
G-PCC does.

\begin{table}[!ht]
\centering
\caption{Six-sequence BD-rate (\%) of concatenation against per-frame
coding at GoF-$30$, with the concatenated branch charged no frame index,
$4.91$ bits per Gaussian (as in Fig.~\ref{fig:avg-grid}), or $8$ bits.}
\label{tab:app-note-recovery}
\footnotesize
\setlength{\tabcolsep}{4pt}
\begin{tabular}{@{}llrrr@{}}
\toprule
Codec & Frame index & Tracked & Semi-tracked & Untracked\\
\midrule
L-GSC     & not charged & $-74.7$ & $-67.7$ & $+2.9$\\
          & $4.91$ bits & $-71.8$ & $-63.4$ & $+5.0$\\
          & $8$ bits    & $-70.0$ & $-60.8$ & $+6.3$\\
\addlinespace
LGSCV     & not charged & $-46.7$ & $-54.6$ & $-7.2$\\
          & $4.91$ bits & $-42.0$ & $-49.2$ & $-1.4$\\
          & $8$ bits    & $-39.0$ & $-45.9$ & $+2.2$\\
\addlinespace
GSCodec-S & not charged & $-51.4$ & $-49.0$ & $-7.1$\\
          & $4.91$ bits & $-47.5$ & $-45.2$ & $-3.5$\\
          & $8$ bits    & $-45.0$ & $-42.8$ & $-1.2$\\
\bottomrule
\end{tabular}
\end{table}

Without the charge, the tracked range of Sec.~\ref{sec:exp1} would be
$-43.0\%$ (SPZ) to $-74.7\%$ (L-GSC), and the untracked gains of LGSCV and
GSCodec-S $-7.2\%$ and $-7.1\%$; at $8$ bits, the latter become $+2.2\%$
and $-1.2\%$. The codec rankings of Sec.~\ref{sec:exp1} and the
semi-tracked range, $-37.9\%$ (HGSC) to $-72.2\%$ (G-PCC), hold under every
charge, and on tracked and semi-tracked input all three codecs keep gains
of at least $39\%$.

\subsection{P-frames of D-GPCC and D-FCGS}
\label{app:note-pframes}
Concatenated key frame coding replaces only the intra codec of D-FCGS
(Sec.~\ref{sec:experiments}): D-GPCC runs the released D-FCGS motion
network, checkpoint, entropy model and parameters without modification. The
P-frame bitstreams, however, are not identical between the two methods, because the P-frames of each method
are coded against its own decoded I-frames and in the Gaussian order of
those I-frames. On \emph{Flame Steak} at GoF-$10$, the P-frames cost
$0.49$--$0.50$\,MB per frame in D-FCGS, $0.44$\,MB in D-GPCC with its
I-frames coded one by one, and $0.42$--$0.43$\,MB in D-GPCC. On one GoF of
the same sequence, reordering the Gaussians from their original order into
the spatial output order of G-PCC changes the reconstruction by less than
$10^{-5}$\,dB but reduces the P-frame bytes by $8.0\%$, because the position
side channel of the P-frames compresses better in that order. Part of the
$-46.2\%$ of D-GPCC against D-FCGS therefore comes from cheaper P-frames
rather than from concatenation. The effect of concatenation alone is
measured by comparing D-GPCC with and without concatenated I-frames, where
the P-frame codec and the I-frame codec are the same: coding the I-frames
as one concatenated set reduces BD-rate by $27.4$--$35.8\%$ at GoF-$10$ and
by $41.3$--$49.9\%$ at GoF-$5$ across the six sequences.

% ---------------------------------------------------------------------
\section{Sequences and Estimators}
\label{app:sequences}
All experiments use all $300$ frames of the six N3DV sequences
\emph{Coffee Martini}, \emph{Cook Spinach}, \emph{Cut Roasted Beef},
\emph{Flame Salmon}, \emph{Flame Steak} and \emph{Sear Steak}. We train
every estimator with its released code on each sequence and export one
Gaussian set per frame. All frames are stored in the $62$-property layout
of the reference 3DGS implementation at SH degree~$3$; lower SH degrees
are zero-padded per colour channel. Table~\ref{tab:app-estimators}
summarizes the four estimators.

\begin{table}[!ht]
\centering
\caption{Estimators. Gaussian counts are ranges over all frames of the six
sequences; a tracked sequence keeps one count. Export loss is the PSNR
against the captured images of the held-out camera that is lost between
the estimator's own rendering and its exported frame-wise Gaussian sets,
averaged over the six sequences.}
\label{tab:app-estimators}
\footnotesize
\setlength{\tabcolsep}{3pt}
\begin{tabular}{@{}llcc@{}}
\toprule
Estimator & Pattern & \shortstack{Gaussians\\per frame} &
\shortstack{Export\\loss (dB)}\\
\midrule
3DGStream, base stream & tracked & 357k--708k & $0.27+0.80$\\
4DGaussians & tracked & 118k--132k & $2.39$\\
QUEEN & semi-tracked & 248k--445k & $0.88$\\
INRIA 3DGS, per frame & untracked & 281k--845k & ---\\
\bottomrule
\end{tabular}
\end{table}

\emph{3DGStream} starts from a standard 3DGS model and obtains every later
frame by applying a learned transformation to the Gaussians of the
previous frame; it also spawns Gaussians for newly appearing content.
Exporting its state at every frame costs $0.27$\,dB. The base stream keeps
only the Gaussians of the first frame, so the Gaussian count is constant
and the index $n$ identifies the same Gaussian in every frame. This costs
a further $0.80$\,dB but makes the sequence tracked, which D-FCGS, D-GPCC
and GSCodec-D require.

\emph{4DGaussians} deforms one canonical Gaussian set with a learned
deformation field, and its official exporter evaluates the field at each
frame. Writing a continuous deformation as static frames is the most
lossy export of the four ($2.39$\,dB on average, up to $3.9$\,dB), which
is why the main comparison uses 3DGStream as its tracked input.
4DGaussians enters the estimator sweep of App.~\ref{app:perseq}.

\emph{QUEEN} maintains one Gaussian set with gated per-frame updates,
additions and removals; we export its state after each frame
($0.88$\,dB). \emph{INRIA 3DGS} is optimized independently for every frame
from the same calibrated camera poses, so the frames share the scene but
no Gaussians, and the frames themselves are the representation.

Table~\ref{tab:app-conditioning} lists every edit applied to the exported
frames before coding. No other change is made to any input: no pruning,
resampling or reordering, and no per-sequence tuning of a codec. Screening
for non-finite values runs on all $24$ sequences.

\begin{table}[!ht]
\centering
\caption{Every edit applied to the exported frames before coding.}
\label{tab:app-conditioning}
\footnotesize
\setlength{\tabcolsep}{3pt}
\begin{tabularx}{\columnwidth}{@{}Y>{\raggedright\arraybackslash}p{2.35cm}>{\raggedright\arraybackslash}p{1.75cm}@{}}
\toprule
Edit & Sequences & Extent\\
\midrule
SH degree $1\!\to\!3$ zero-padding; quaternion sign canonicalized &
3DGStream, all & layout, sign\\
Non-finite values: $\pm\infty\!\to\!\pm60$, NaN$\,\to\!0$ &
3DGStream, all & $<\!10^{-6}$ of values\\
SH degree $2\!\to\!3$ zero-padding & QUEEN, all & layout\\
Degenerate Gaussians removed (log-scale $>2$) & QUEEN, all &
$\approx\!0.2\%$ of Gaussians\\
Gaussians with non-finite values removed & QUEEN, \emph{Coffee Martini} &
$1$ per frame in $127$ frames\\
Lower G-PCC input scale$^\dagger$ & 4DGaussians, \emph{Coffee Martini},
\emph{Flame Salmon}; QUEEN, \emph{Coffee Martini} & none\\
\bottomrule
\end{tabularx}
\raggedright\footnotesize
$^\dagger$An encoder setting, not an edit of the data: coordinates beyond
$\pm1024$ overflow $32$-bit arithmetic at the default input scale. The
4DGaussians exporter leaves about $0.25\%$ of every frame's Gaussians far
outside the scene, so those two frames span $6827$ and $4328$ units while
$99\%$ of their Gaussians fit inside $120$ and $114$; we keep them, and
App.~\ref{app:perseq} reports what they cost the per-frame branch.
\end{table}

% ---------------------------------------------------------------------
\section{Evaluation Protocol}
\label{app:protocol}

\subsection{Distortion}
\label{app:distortion}
Distortion compares renders of the decoded Gaussian sets with renders of
the input Gaussian sets (Sec.~\ref{sec:experiments}). We call this the
identical-source reference, the name used in the repository and on the
project page. Both are rendered with the same rasterizer (gsplat, black
background) from every calibrated view of the sequence, and squared error
is pooled per colour plane over all pixels, views and frames before it is
converted:
\begin{equation}
\mathrm{PSNR}_{\mathrm{RGB}}=10\log_{10}\frac{1}
{\tfrac13\left(\mathrm{MSE}_R+\mathrm{MSE}_G+\mathrm{MSE}_B\right)},
\label{eq:app-psnr}
\end{equation}
with pixel values in $[0,1]$. $\mathrm{PSNR}_{\mathrm{YUV}}$ is computed in
the same way from BT.709 full-range $4{:}4{:}4$ planes with $1{:}1{:}1$
weights. All BD-rates in the paper use $\mathrm{PSNR}_{\mathrm{RGB}}$; with
$\mathrm{PSNR}_{\mathrm{YUV}}$, none of the $18$ six-sequence BD-rates of
Fig.~\ref{fig:avg-grid} changes by more than $0.6$ percentage points.

A result is accepted only if all frames are scored, pooled
$\mathrm{PSNR}_{\mathrm{RGB}}$ is at least $15$\,dB, and the average of
per-frame decibels exceeds the pooled value by less than $3$\,dB. The last
two conditions detect a collapsed sequence and damage confined to a few
frames or views. Every reported result passes.

\begin{figure}[!ht]
\centering
\includegraphics[width=\columnwidth]{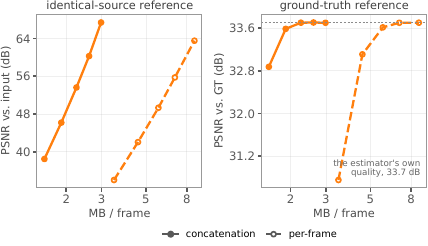}
\caption{The same four runs of G-PCC on 3DGStream \emph{Flame Steak},
scored against the two references. Against the input, the rate points span
$30$\,dB and the two branches stay apart, so a BD-rate between them is
well posed. Against the held-out camera, both branches run into the quality
of the estimator itself and flatten within a few tenths of a decibel of
each other, which leaves almost no quality interval to integrate over.}
\label{fig:app-reference}
\end{figure}

We also render the decoded frames at the held-out camera and compare them
with the captured images (PSNR, SSIM, and LPIPS with AlexNet features,
averaged over frames). These ground-truth curves are on the project page
but enter no BD-rate: both branches saturate at the quality of the estimator,
so the quality interval they share collapses
(Fig.~\ref{fig:app-reference}), and a single view does not
see damage outside its field of view. For example, one QUEEN sequence
carried a non-finite Gaussian center in each of its last $127$ frames.
Before this was removed, the ground-truth PSNR of its G-PCC concatenated
branch was $24.88$\,dB, and $24.90$\,dB after; the identical-source PSNR rose
from $39.7$ to $61.9$\,dB.

\subsection{Rate}
Rate is the total size of all bitstreams and side files read by the
codec's decoder, divided by the number of frames, in MB per frame
($1$\,MB$\;=10^6$ bytes). A concatenated GoF is one coding unit, and its
bytes are counted once. How the frame of each decoded Gaussian is
recovered differs between codecs (App.~\ref{app:recovery}); for L-GSC,
LGSCV and GSCodec-S, which do not transmit it, the rate of the concatenated
branch adds the cost of a $\log_2 30$-bit index per Gaussian
(App.~\ref{app:note-recovery}). For
D-FCGS and D-GPCC, the rate includes all I-frame and P-frame bitstreams.

The HGSC rates were corrected once after scoring. A restarted task
appended a second, identical ledger row for a frame, and two tasks writing
to one ledger at the same time lost a row. Both errors favoured
concatenation. Rates now come from exactly one row per frame or GoF,
checked to cover the whole sequence, and the reported six-sequence
BD-rates ($-66.4\%$ tracked, $+27.8\%$ untracked) are the corrected ones.

\subsection{BD-rate and six-sequence averages}
BD-rate follows~\cite{bjontegaard2001}: the logarithm of the rate of each
curve is fitted as a cubic polynomial of PSNR (a quadratic for the three
rate points of L-GSC), and both fits are integrated over the PSNR interval
that the two curves share. A negative value means that concatenation needs
less rate for the same quality. A six-sequence average curve averages the
rate and the PSNR of the $i$-th rate point, sorted by rate, over the six
sequences, and the average BD-rate is computed between two average curves.
It is therefore not the mean of the per-sequence BD-rates in
App.~\ref{app:perseq}.

\subsection{Throughput, memory, and delay}
In Table~\ref{tab:gof-full}, encoding time (Enc\,T) runs from the input
frames to the last bitstream byte and includes concatenation and
partitioning; decoding time (Dec\,T) runs from the bitstreams to per-frame
Gaussian sets and includes splitting the GoF. Both are divided by the
$300$ frames and averaged over the rate points.

The measurement is one job. A group length is not comparable with another
unless both met the same machine and the same amount of competition for it,
and the number of slices a GoF is cut into grows with the GoF, so any fixed
number of parallel coding processes would give the longer groups more
company than the shorter ones. Every rate point therefore runs all six group
lengths back to back inside one task, on one node, with one coding process
at a time and one thread in every library, and the six lengths run in a
different order in each of the five tasks, so a node that drifts over the
hours of a task cannot drift along the group-length axis. A fixed
single-frame job between every pair of measurements reports what the node
was doing: within a task those probes scatter by $11$--$21\%$ without a
trend, while the group-length effect is monotone.

Memory is the peak resident set of the coder's own process tree, sampled
every $0.1$\,s. It is not the figure a job scheduler reports: that one also
counts the page cache of every PLY the run reads and writes, which on this
pipeline is the larger half of it and is reclaimed before any process is
killed, so it says more about how much memory the job was granted than about
how much it needed. The two columns separate the only part that grows from
the part that does not. \emph{Concat} is the peak of the pass that
concatenates the GoF and derives its slice partition: it holds the whole
group and costs $22.3$\,MB per frame, linear from $0.065$\,GB at GoF-$1$ to
$3.38$\,GB at GoF-$150$. \emph{Encode} is the peak of the coder itself
and is bounded instead: the partition caps a slice at $1.1$M Gaussians, and
the peak, which falls on the decoder rather than the encoder, is
$3.3$\,GB per million Gaussians of the largest slice, so it rises only while
the slices do and settles at $1.29 \to 3.75$\,GB.
Per-frame coding sits below that bound, not on it, because one frame of this
sequence is $357$k Gaussians and is never split.

\begin{figure}[!ht]
\centering
\includegraphics[width=\columnwidth]{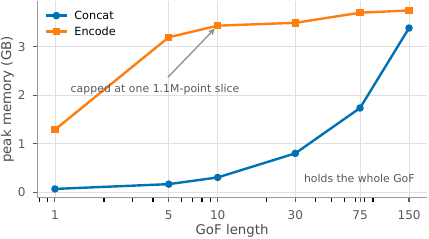}
\caption{The two memory columns of Table~\ref{tab:gof-full}. Only
\emph{Concat} grows with the group; \emph{Encode} is bounded by the slice
budget and stops. The two meet at GoF-$150$, where a slice first reaches the
cap.}
\label{fig:app-mem-gof}
\end{figure}

Delay is the structural delay of collecting a GoF at $30$ frames per second.
Our unoptimized implementation adds processing time on top of it ($294$\,s
before the first frame of a GoF-$30$ set is available), which is not
included.

% ---------------------------------------------------------------------
\section{Static GS Codecs}
\label{app:codecs}
Table~\ref{tab:app-codecs} lists the six static GS codecs, their rate
points, and how each returns decoded Gaussians to their frames. Both branches
of a codec use identical settings at every rate point, and every rate
point codes the same Gaussians. The importance-based pruning of HGSC,
enabled by default at $60\%$, is disabled. LGSCV maps Gaussians to square
images and therefore keeps the $\lfloor\sqrt{N}\rfloor^2$ Gaussians of
highest importance; on the $10.70$M Gaussians of a GoF-$30$ set of
3DGStream \emph{Flame Steak}, this removes $0.43\%$, and the same rule
applies in the per-frame branch. GSCodec-S pads its square images with
Gaussians of negligible opacity; the per-frame branch keeps them in the
decoded frames, as the codec defines, and the concatenated branch removes
them when splitting. The SPZ decoder writes an opacity of one as an
infinite logit, which we clamp to $\pm12$ before rendering in both branches.

\begin{table*}[t]
\centering
\caption{Static GS codecs. Rate points are swept identically in both branches.
Frame recovery states how each decoded Gaussian of a concatenated set is
assigned to its frame; $^\ast$marks encoder-side information that is not
transmitted; the rate is charged $\log_2 30$ bits per Gaussian in its place
(Apps.~\ref{app:note-recovery} and~\ref{app:recovery}).}
\label{tab:app-codecs}
\footnotesize
\setlength{\tabcolsep}{4pt}
\begin{tabularx}{\textwidth}{@{}llY>{\raggedright\arraybackslash}p{3.4cm}>{\raggedright\arraybackslash}p{4.3cm}@{}}
\toprule
Codec & Family & Coding & Rate points & Frame recovery\\
\midrule
G-PCC & geometry & TMC13 v23.0-rc2 with our Gaussian splat coding
profile: octree geometry, RAHT for SH, predictive transform for opacity,
scale and rotation & 5: four attribute QP settings, attributes at unit
step & frame index coded losslessly as an $8$-bit attribute\\
HGSC & geometry & G-PCC geometry, anchors by farthest-point sampling, two
levels of detail predicted from coded neighbours & 4: position
$\mathrm{qp}\in\{18,\dots,15\}$ with LoD bit depths & frame index coded
losslessly as an $8$-bit attribute of its geometry\\
L-GSC & geometry & fixed bit-width quantization, Morton ordering, zlib &
3: bit-width presets & Morton sort of the encoder replayed on the input
positions$^\ast$\\
SPZ & geometry & per-Gaussian quantization, zstd & 5: SH bit depths &
order preserved; split by per-frame counts\\
LGSCV & video & Morton ordering and MiniPLAS into 2D maps, SH reduced by
PCA, HEVC & 5: released QP settings & map permutation recorded at the
encoder$^\ast$\\
GSCodec-S & video & PLAS sorting into 2D maps, all-intra HEVC & 4:
released QP settings & sort permutation recorded at the encoder$^\ast$\\
\bottomrule
\end{tabularx}
\end{table*}

Only G-PCC spatially partitions the concatenated set. Its encoder divides
the set along the longest axis of its bounding box into slices of at most
$1.1$M Gaussians, and every slice is coded as an independent unit with its
own coordinate scaling. HGSC applies the same bound when it codes its
anchors with G-PCC. The other codecs code a GoF-$30$ set as a single unit;
the largest sets, from the two QUEEN sequences with $437$k Gaussians per
frame, need about $80$\,GB of memory in HGSC and $52$\,GB in LGSCV.

\subsection{Recovering frames after decoding}
\label{app:recovery}
A concatenated set is decoded as one Gaussian set, and each Gaussian must
then be assigned to its frame. Geometry cannot do this, because the frames
of one GoF are nearly co-located, which is precisely what concatenation
exploits: on the uncompressed GoF-$30$ set of 3DGStream \emph{Flame Steak},
only $66\%$ of nearest-neighbour queries return a Gaussian of the correct
frame. The six codecs recover frames in three ways.

G-PCC and HGSC transmit the frame index. It is an $8$-bit attribute coded
losslessly, spread over the code range in steps of $255/(L-1)$ so that the
attribute scaling of the encoder cannot merge indices, and its cost is part
of the rate (App.~\ref{app:frameindex}). In HGSC, merging of duplicate
positions is disabled, so co-located Gaussians of different frames are not
fused. The per-frame branch runs the same configuration with a constant
index. SPZ preserves the order and number of Gaussians, so a decoded set is
split by the per-frame counts.

L-GSC, LGSCV and GSCodec-S reorder the Gaussians (by Morton code, by Morton
code followed by MiniPLAS, and by PLAS), and none of them has a channel
that could carry a per-Gaussian index. To evaluate concatenation in these
codecs without redesigning them, we recover the frame of each decoded
Gaussian from the reordering computed at the encoder: the Morton sort of
L-GSC is replayed on the input positions, and the permutations of LGSCV
and GSCodec-S are recorded when their maps are built. For L-GSC, we also
verify that every decoded Gaussian lies within quantization error of the
source Gaussian it is assigned to. This information is not transmitted; in its
place, the rates of these three codecs are charged a uniformly coded index
(App.~\ref{app:note-recovery}).

\subsection{Keeping Gaussian identity}
Concatenation exposes faults that per-frame coding does not reach, because
Gaussians of different frames coincide or nearly coincide. In HGSC as
released, farthest-point sampling returned the same Gaussian repeatedly
once the distinct positions of a block ran out ($31\%$ of the selection was
duplicated on a GoF-$30$ set), and the anchor coder returned anchors in
sorted order while their frame indices remained in sampling order, which
assigned $7.7\%$ of the reconstruction to wrong frames. Both faults caused
a quality ceiling that no rate setting could lift ($0.2$\,dB over a $36\%$
range of rate), and both are corrected in all reported results. On QUEEN
input, whose persisting Gaussians keep identical positions, the recursive
KD-tree split of HGSC also does not terminate on a block of identical
positions; such blocks are split into chunks of the leaf size. The
repository lists every modification of this kind for all codecs.

% ---------------------------------------------------------------------
\section{Inter-Frame Similarity}
\label{app:similarity}
The matrices of Fig.~\ref{fig:wd} evaluate the cost of
Eqs.~\eqref{eq:w2}--\eqref{eq:framecost} with the level-of-detail structure of
G-PCC's predicting transform and the quantization steps of our G-PCC
configuration, so that $D(t,u)$ approximates what G-PCC would pay to predict the
Gaussians of one frame from those of the other. It differs from the conceptual
form of Sec.~\ref{sec:correlation_pattern} as follows.
\begin{enumerate}[leftmargin=*,nosep]
\item \emph{Neighbourhood.} The two frames are concatenated and their positions
quantized on one grid of $2^{18}$ steps spanning the pair, the coding grid of
G-PCC. A level-of-detail pass as in G-PCC's predicting transform keeps one
Gaussian per occupied voxel at each level, with the voxel
starting at the median nearest-neighbour distance, at least two grid steps, and
doubling at each level. Every Gaussian the pass removes is compared with a
single prediction, the inverse-distance-weighted average of its three nearest
kept Gaussians, which may come from either frame, rather than with each of its
three nearest Gaussians in the other frame as in Eq.~\eqref{eq:framecost}.
\item \emph{Shape term.} $W_2$ enters unsquared and only through the
covariances: the term is $W_2(\mathbf{\Sigma}_i,\hat{\mathbf{\Sigma}}_i)$
between the covariance of the Gaussian and the inverse-distance-weighted mean of
the covariances of its neighbours, divided by the square root of the mean
covariance trace of the sequence. Positions act through the neighbour search and
its weights rather than as the term
$\lVert\boldsymbol{\mu}_i-\boldsymbol{\mu}_j\rVert^2$ of Eq.~\eqref{eq:w2}.
\item \emph{Appearance terms.} In place of the weights $\beta$ and $\gamma$ of
Eq.~\eqref{eq:wdcost}, the SH DC term, the SH bands of degree $1$, $2$ and $3$
and the opacity each contribute the RMS of their residual, divided by the
standard deviation of that attribute over the sequence and weighted by the
relative quantization step $2^{-(\mathrm{QP}_b-\mathrm{QP}_0)/6}$ of the lowest
G-PCC rate point (QPs $26$, $38$, $46$, $49$ and $46$, i.e., weights $1$,
$0.25$, $0.10$, $0.07$ and $0.10$). The normalization makes the shape and
appearance terms comparable without fixing a rate point, which is what
$\alpha$, $\beta$ and $\gamma$ stand for in the main text.
\item \emph{Averaging.} $D(t,u)$ is the mean cost over all Gaussians of the
pair, with the Gaussians the pass keeps costing zero, so the normalization is
$N_t+N_u$ rather than $3N_t$.
\item \emph{Sampling and numerics.} Each frame is first reduced to $250$k
Gaussians by the same octree-centroid rule for every estimator. Non-finite
values are replaced by the mean of their attribute, missing SH coefficients are
set to zero, covariances are rebuilt from log-scales and quaternions, and $W_2$
is computed exactly from eigendecompositions.
\item \emph{Normalization of the matrix.} $\widetilde{D}$ is min--max
normalized over the off-diagonal entries of each matrix separately, because
absolute cost levels differ between estimators.
\end{enumerate}
The inputs are the exported frames before the edits of
Table~\ref{tab:app-conditioning}; for 3DGStream, every frame of \emph{Flame
Steak} has the same Gaussian count as the base stream used for coding,
$356{,}663$. As a scalar summary we use
\begin{equation}
\rho=\frac{\operatorname{mean}_{|t-u|>150}D(t,u)}
          {\operatorname{mean}_{|t-u|=1}D(t,u)},
\label{eq:app-rho}
\end{equation}
over all $44{,}850$ pairs of frames, which equals one when distant frames
are as similar as adjacent ones. Table~\ref{tab:app-rho} lists $\rho$ for
all four estimators on \emph{Flame Steak}. Untracked input stands out with
$\rho\approx1$. Among the other estimators, $\rho$ does not rank the gains:
4DGaussians has the lowest $\rho$ of the three but the largest G-PCC gains
in the estimator sweep (Table~\ref{tab:app-sweep}).

\begin{table}[!ht]
\centering
\caption{Temporal structure of \emph{Flame Steak} for the four estimators,
over all $300$ frames.}
\label{tab:app-rho}
\footnotesize
\begin{tabular}{@{}llcc@{}}
\toprule
Estimator & Pattern & $\rho$ & Range of $D$\\
\midrule
QUEEN & semi-tracked & $1.571$ & $0.362$--$1.046$\\
3DGStream & tracked & $1.292$ & $0.776$--$1.087$\\
4DGaussians & tracked & $1.179$ & $0.432$--$0.552$\\
INRIA 3DGS & untracked & $1.002$ & $1.295$--$1.442$\\
\bottomrule
\end{tabular}
\end{table}

% ---------------------------------------------------------------------
\section{Mechanisms Behind the Gains}
\label{app:mechanisms}

\subsection{Spatial overlap}
The overlap figures of Secs.~\ref{sec:exp1} and~\ref{sec:exp5} quantize
the positions of the GoF-$30$ set of \emph{Flame Steak} with the position
quantizer of HGSC at its finest setting, $\mathrm{qp}{=}18$, and count the
Gaussians that share a voxel with at least one other Gaussian: $33.7\%$ for
3DGStream, $89.2\%$ for QUEEN and $1.0\%$ for INRIA 3DGS.

Semi-tracked input overlaps more than tracked input because a Gaussian
that QUEEN keeps also keeps its exact position. Between adjacent frames,
$90.5$--$92.8\%$ of the Gaussians of QUEEN do not move (only $4.6\%$
between the first two frames, where the initial reconstruction is
replaced), whereas 3DGStream moves about two thirds of its Gaussians slightly in
every frame (App.~\ref{app:tracked}).
In a GoF-$30$ set of \emph{Cook Spinach}, $14.1\%$ of the positions of
QUEEN are distinct, against $63.1\%$ for 3DGStream. The number of occupied
voxels of the set, relative to the sum over its frames, stays at $11.4\%$
for QUEEN from $\mathrm{qp}{=}15$ to $18$, while it grows from $30.9\%$ to
$62.8\%$ for 3DGStream: a finer grid separates slightly moved copies but
not identical ones. Accordingly, G-PCC, LGSCV and SPZ gain more on
semi-tracked than on tracked input, by $5.3$, $7.9$ and $17.5$ percentage
points (Fig.~\ref{fig:avg-grid}).

\subsection{Tracked input}
\label{app:tracked}
3DGStream moves its Gaussians but keeps their appearance. Over frames $0$
to $29$ of \emph{Flame Steak}, all $356\,663$ Gaussians keep their SH
coefficients bit for bit; from frame $1$ on, $96.0\%$ also keep their
opacity and $98.3\%$ their scale, which change once between frames $0$ and
$1$. $34.8\%$ keep their position, and the others move by a median of
$6.6\cdot10^{-4}$, while no Gaussian keeps its rotation. A concatenated set
thus holds $30$ nearly co-located copies of each Gaussian with the same
appearance, and every codec gains on every tracked sequence
(Table~\ref{tab:app-perseq}). How much depends on whether the codec places
the copies next to each other and predicts one from another.

\begin{figure}[!ht]
\centering
\includegraphics[width=\columnwidth]{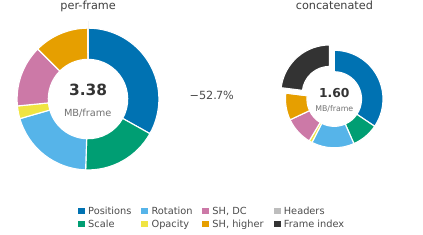}
\caption{The same $30$ frames of 3DGStream \emph{Flame Steak}, coded one by
one and coded as one set, by G-PCC at its coarsest rate point. Ring areas
are proportional to the totals in their centres. Every stream shrinks; the
frame index is the one the concatenated run adds.}
\label{fig:app-tracked-gpcc}
\end{figure}

\emph{Prediction from neighbours (G-PCC, $-66.9\%$; HGSC, $-66.4\%$).} Both
codecs predict attributes from nearby Gaussians, and in a concatenated set
the nearest ones are copies from other frames.
Fig.~\ref{fig:app-tracked-gpcc} breaks G-PCC down on the first GoF-$30$
set of \emph{Flame Steak} at its coarsest rate point. Every attribute
stream shrinks, by $51\%$ for positions to $83\%$ for opacity, including
rotation, which changes in every frame but only slightly, so its residuals
stay small. Without the frame index, the set would cost $64\%$ less than
the per-frame branch; with it, $53\%$ less. HGSC gains mostly in its detail
stream, which predicts each detail Gaussian from its anchors: at its finest
rate point, this stream costs $3.07$ against $12.23$\,MB per frame, while
geometry, including the frame index, goes from $1.23$ to $0.95$\,MB and the
anchors from $1.30$ to $0.87$\,MB. The same predictor gains less on QUEEN,
whose attributes change slightly in every frame (App.~\ref{app:semitracked}).

\emph{Ordering (L-GSC, $-71.8\%$; SPZ, $-43.0\%$).} Neither codec predicts
across Gaussians: both quantize each Gaussian and compress the resulting
bytes losslessly, so the gain depends on how close repeated values end up.
L-GSC sorts the set by Morton code with a stable sort, so the $30$ copies of
each static Gaussian share a code and follow each other in frame order,
with identical SH coefficients. SPZ keeps the input order, so the next copy
of a Gaussian lies one frame, $356\,663$ Gaussians, further on in each
attribute stream, where only the long-range matching of zstd can find it.

\emph{Video-based codecs (LGSCV, $-42.0\%$; GSCodec-S, $-47.5\%$).} Sorting
also places the copies next to each other in the 2D maps, which HEVC codes
lossily. Their frame index is charged separately and takes $5.7$--$10.3\%$
and $5.9$--$9.6\%$ of their rates on tracked input
(App.~\ref{app:note-recovery}). The rates of the individual maps were not
recorded, so we do not break these gains down further.

\subsection{Semi-tracked input}
\label{app:semitracked}
\emph{HGSC} gains much less on semi-tracked input ($-37.9\%$) than on
tracked input ($-66.4\%$). Its predictor does find the right Gaussian:
$91\%$ of the detail Gaussians of a concatenated set take their nearest
anchor from a co-located copy in another frame, against $0.5\%$ with
per-frame coding. But QUEEN re-optimizes the attributes of persisting
Gaussians in every frame, so the residuals are small but never zero:
between adjacent frames, the median change is $0.001$ for the SH DC term
and $0.0006$--$0.0009$ for the higher SH bands, whereas 3DGStream leaves
them unchanged for all Gaussians. HGSC quantizes residuals over their own
range at a fixed bit depth and compresses the packed integers with zlib.
At coarse bit depths, this jitter falls within one quantization step and
costs nothing; at fine bit depths, it becomes low-order noise that does not
compress. On \emph{Cook Spinach}, the geometry and anchor streams of the
concatenated branch are cheaper at every rate point ($0.41$ against
$0.90$\,MB and $0.92$ against $1.54$\,MB per frame at the finest), but its
detail stream grows from $0.88$ to $10.76$\,MB per frame across the rate
points, against $2.63$ to $6.65$\,MB for the per-frame branch, and the two branches
cross at the second finest point. Only $0.3\%$ of the SH residuals in the
concatenated set are exactly zero, against $60\%$ within a frame.

\emph{SPZ} gains most on semi-tracked input for a reason specific to QUEEN,
whose higher SH coefficients come from a codebook: a frame of \emph{Cook
Spinach} holds $233$ distinct higher-order SH vectors for $261$k Gaussians,
whereas nearly every Gaussian of 3DGStream and INRIA 3DGS has its own. The
rate points of SPZ only change the SH precision, so on QUEEN input they
change quality but hardly rate (per-frame branch: $25.0$ to $50.7$\,dB within
$0.15$\,MB per frame). The curves are nearly vertical, and BD-rate
essentially measures the rate ratio of the two branches.

\subsection{Untracked input}
\label{app:untracked}
With $1.0\%$ of Gaussians sharing a voxel, concatenation adds Gaussians
without adding coincident positions, and three mechanisms determine the
result of each codec.

\emph{Quantization steps derived from the data (HGSC, $+27.8\%$).} HGSC
takes both of its steps from the range of the set it is given: positions
are quantized over the bounding box of the set at its position setting, and
the two levels of detail are coded as residuals with respect to the nearest
Gaussian already coded, quantized over their own range at the bit depth of
the rate point. Concatenation widens both ranges on untracked input. The
bounding box grows by the factor above. The residuals grow because HGSC
splits the set into KD-tree leaves of at most $256$ Gaussians and promotes
a tenth of each leaf to anchors: with $30$ times more Gaussians in the same
volume, the nearest already-coded Gaussian usually comes from another
frame, and independently estimated frames reach different local optima, so
its attributes differ even where the rendered surface agrees. Both effects
are distortion that no rate point buys back, and that is what the curves
show: on \emph{Flame Steak}, the concatenated branch is slightly cheaper than
the per-frame branch at every rate point ($28.9$ against $29.4$\,MB per frame
at the finest, $16.2$ against $17.2$ at the second coarsest) and $3$ to
$9$\,dB worse ($51.8$ against $60.7$\,dB, $41.7$ against $50.9$\,dB).
Duplicate merging is disabled in both branches, so this is not a collision; per
sequence, the loss ranges from $+18.3\%$ to $+50.4\%$.

\begin{figure}[!ht]
\centering
\includegraphics[width=\columnwidth]{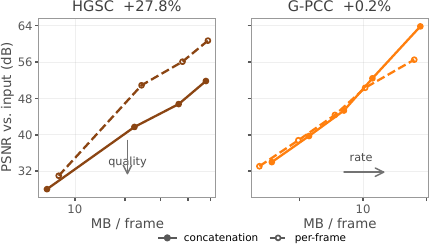}
\caption{The two failures have different shapes on untracked \emph{Flame
Steak}. Concatenation moves HGSC down, spending quality no rate point buys
back, and moves G-PCC right, spending only rate. Both panels share a
vertical axis.}
\label{fig:app-untracked-shape}
\end{figure}

\begin{figure}[!ht]
\centering
\includegraphics[width=\columnwidth]{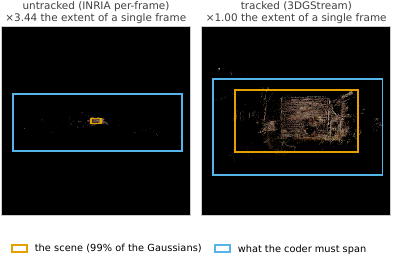}
\caption{Every Gaussian of the first GoF-$30$ group of \emph{Flame Steak}
as one point, with no opacity: it is the positions that set a coder's range.
The scene is nearly the same size in both panels, $52$ and $49$ world units
across, but it covers $6\%$ of the range the untracked group makes a coder
span and $73\%$ of the tracked one. The two panels are at different
scales.}
\label{fig:app-spatial}
\end{figure}

\emph{Wider ranges, absolute steps (L-GSC, G-PCC, SPZ).} Every
independently estimated frame places its outlying Gaussians differently, so
the per-axis extent of the concatenated set is on average $3.44$ times that
of a single frame, against $1.00$ for 3DGStream and QUEEN
(Fig.~\ref{fig:app-spatial}). The widening is the work of a small minority:
a box $52$ world units across holds $99\%$ of the untracked group and one of
$49$ holds $99\%$ of the tracked one, but that box is $6\%$ of the range the
untracked group makes a coder span and $73\%$ of the tracked one. L-GSC quantizes
to fixed bit widths over this range and pays with distortion, as HGSC does:
on tracked input its two branches reach the same PSNR at every rate point, but
on untracked \emph{Flame Steak} the concatenated branch reaches $37.00$\,dB
against $38.17$\,dB at the coarsest point. Per sequence, the result of L-GSC
ranges from $-1.6\%$ to $+10.1\%$, depending on the outliers of each scene.
G-PCC is the counter-example that separates the mechanism from the
prediction: it also predicts attributes from neighbours (opacity, scale and
rotation through the predictive transform; the SH coefficients are coded by
RAHT without prediction), but its attribute steps are absolute quantization
parameters and its geometry is coded losslessly on a grid derived per
slice, so a wider set and worse predictions cost rate, not quality. On
untracked \emph{Flame Steak} its concatenated branch needs $7$ to $15\%$ more
rate per point at equal or better quality ($3.69$ against $3.22$\,MB per
frame at $34.0$ against $33.1$\,dB), which integrates to $+0.2\%$. SPZ
quantizes each Gaussian independently, so both branches have identical
distortion, and it finds little to match across frames ($-0.2\%$).

Table~\ref{tab:app-hgsc-gpcc} puts the two geometry-based codecs side by
side, because their results, $+27.8\%$ and $+0.2\%$, are often read as a
difference in prediction. Both quantize positions, both predict attributes
from neighbours, and both see the same widened set. What separates them is
where their quantization steps come from. HGSC derives every step from the
data it is handed, so a wider bounding box and wider residuals mean coarser
steps at the same bit depth, and the loss lands on quality: on untracked
\emph{Flame Steak} its concatenated branch is $0.5$\,MB per frame cheaper than
its per-frame branch and $8.9$\,dB worse. G-PCC quantizes attributes with
absolute parameters and codes geometry losslessly on a grid derived per
slice, so the same widening leaves quality alone and is paid in rate:
$0.5$\,MB per frame more at $0.9$\,dB better. A codec whose steps follow the
data can therefore lose on untracked input however good its predictor is,
while one with absolute steps stays near its baseline however poor its
predictor becomes.

\begin{table}[!ht]
\centering
\caption{Why concatenation costs HGSC quality and G-PCC only rate on
untracked input. The last two rows compare the concatenated with the
per-frame branch at the finest rate point of each codec, on \emph{Flame
Steak}.}
\label{tab:app-hgsc-gpcc}
\footnotesize
\setlength{\tabcolsep}{3pt}
\begin{tabularx}{\columnwidth}{@{}lYY@{}}
\toprule
& HGSC & G-PCC\\
\midrule
Position step & bounding box of the coded set $/\,2^{\mathrm{qp}}$ &
lossless on a grid derived per slice\\
Attribute step & residual range $/\,2^{\mathrm{bit\ depth}}$ &
absolute QP per attribute\\
Attribute predictor & nearest already-coded Gaussian &
RAHT for SH (no prediction); three candidates for opacity, scale, rotation\\
\addlinespace
Rate & $-0.5$\,MB per frame & $+0.5$\,MB per frame\\
Quality & $-8.9$\,dB & $+0.9$\,dB\\
BD-rate & $+27.8\%$ & $+0.2\%$\\
\bottomrule
\end{tabularx}
\end{table}

\emph{Sorting with spatial prediction (LGSCV, $-7.2\%$; GSCodec-S,
$-7.1\%$ before the frame-index charge).} Both video-based codecs sort the concatenated set into 2D maps
and code them with HEVC intra prediction, which benefits whenever
neighbouring map positions hold similar values, regardless of their
frames. Sorting alone is not sufficient: L-GSC also orders Gaussians by
Morton code but has no spatial prediction, and does not gain. Most of
these gains is used up by the frame index they are charged, which leaves
$-1.4\%$ and $-3.5\%$ (App.~\ref{app:note-recovery}).

% ---------------------------------------------------------------------
\section{Frame-Index Overhead}
\label{app:frameindex}
G-PCC and HGSC code the frame index with the same tool: an $8$-bit
attribute, spread over the code range in steps of $255/(L-1)$ and coded by
the predicting transform at unit step, i.e., losslessly. G-PCC codes it on
its lossless geometry with the attribute settings of our splat profile.
HGSC attaches it to the G-PCC call that codes its quantized positions, with
the default prediction settings of its own G-PCC build; this call is
separate from the $16$-bit reflectance channels in which HGSC codes the
Gaussian attributes. Table~\ref{tab:app-frameindex} lists the cost on the
first GoF-$30$ set of \emph{Flame Steak}. For G-PCC, it is the sum over all
slices of the index size that the encoder reports: $9\,055\,243$ bytes for
the $9.38$M Gaussians of INRIA 3DGS, $7\,186\,599$ bytes for the $7.44$M of
QUEEN and $11\,053\,967$ bytes for the $10.70$M of 3DGStream, identically at
the four lossy rate points; the lossless-attribute point codes the index
with another geometry configuration and differs by less than $1\%$. For
HGSC, the tracked values come from the encoder logs of six runs, which
cover its four rate points. All other values come from re-running the
coding of the first group at the coarsest and at the lossless-attribute
rate point, which reproduces the size of the original bitstreams byte for
byte; at the other lossy rate points, G-PCC codes the index identically. Shares are taken within the same group.

\begin{figure}[!ht]
\centering
\includegraphics[width=\columnwidth]{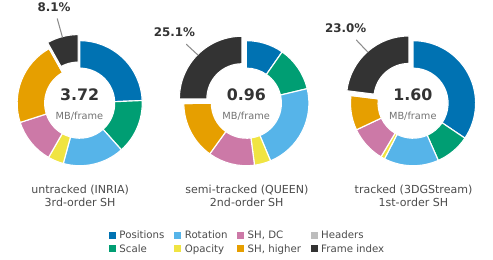}
\caption{Where the rate of a concatenated G-PCC group goes, on the first
GoF-$30$ group of \emph{Flame Steak} at the coarsest rate point. The rings
carry the same streams in the same order and colour, so a wedge can be read
across the three inputs; the centre gives the total the shares are taken
of. The frame index is the only stream that is not scene content, and is
the only wedge pulled out. Slice headers are under $0.02\%$ and not
visible.}
\label{fig:app-rings}
\end{figure}

Both codecs spend about the raw $8$ bits or more, far above the
$\log_2 30=4.91$ bits of an index coded at its uniform entropy, the charge
of L-GSC, LGSCV and GSCodec-S (App.~\ref{app:note-recovery}), because
neighbouring Gaussians in a concatenated set often belong to other frames
and prediction does not help. HGSC spends $1.3$--$2.1$ bits more than
G-PCC; we did not trace this to a specific setting. The share of the index differs between inputs far more than its cost
does, because it is the rest of the bitstream that changes: at the coarsest
rate point, a concatenated G-PCC group costs $3.72$\,MB per frame on
untracked input, whose Gaussians carry third-order SH, against $1.60$ on
3DGStream, which carries only first-order SH, and $0.96$ on QUEEN, which
carries second-order SH; the index itself stays between $0.24$ and
$0.37$\,MB per frame. Fig.~\ref{fig:app-rings} shows what the rest is made
of. Untracked input spends $33.6\%$ of its rate on SH and $24.3\%$ on
positions; on tracked input the appearance streams collapse, because the
$30$ copies of a Gaussian carry the same SH coefficients bit for bit
(App.~\ref{app:tracked}), so SH falls to $18.4\%$ and positions, which
every copy still needs, rise to $34.4\%$ of a bitstream less than half the
size. The index is thus largest in share exactly where concatenation works
best. On untracked input it is at most $4.5\%$ of HGSC's bitstream, far
too little to explain the $+27.8\%$ of HGSC (App.~\ref{app:untracked}).

\begin{figure}[!ht]
\centering
\includegraphics[width=\columnwidth]{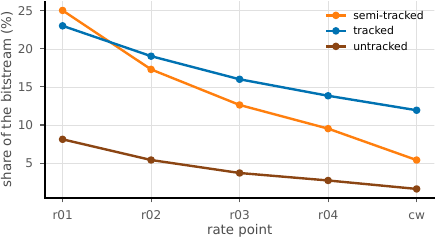}
\caption{The index costs the same bytes at every lossy rate point, so its
share falls as the rest of the bitstream grows, and sits highest on the
input whose Gaussians carry the fewest SH coefficients.}
\label{fig:app-index-share}
\end{figure}

\begin{table*}[t]
\centering
\caption{What the frame index costs every codec, on \emph{Flame Steak} at
GoF-$30$, over the rate points of each codec. G-PCC and HGSC carry the index
inside their bitstreams, so their two rate columns are measured on the first
group and the share is exact within it; the three codecs that reorder their
Gaussians are charged a separate index stream at $\log_2 30$ bits each
(App.~\ref{app:note-recovery}), and their columns are the whole sequence.
SPZ preserves the order and the count, so it needs none. A range is over the
rate points; a single value means the column does not move with the rate.}
\label{tab:app-frameindex}
\footnotesize
\setlength{\tabcolsep}{6pt}
\begin{tabular}{@{}lllrrrr@{}}
\toprule
Codec & Index & Input & Total (MB/f) & Index (MB/f) & Share (\%) & Bits/Gaussian\\
\midrule
G-PCC      & in the bitstream  & tracked       &     $1.60$--$3.08$ &              $0.368$ &     $11.9$--$23.0$ &           $8.26$\\
           &                   & semi-tracked  &     $0.96$--$4.48$ &     $0.240$--$0.242$ &      $5.4$--$25.1$ &   $7.72$--$7.79$\\
           &                   & untracked     &    $3.72$--$18.72$ &     $0.300$--$0.302$ &       $1.6$--$8.1$ &   $7.67$--$7.72$\\
\midrule
HGSC       & in the bitstream  & tracked       &     $2.26$--$4.88$ &     $0.437$--$0.439$ &      $9.0$--$19.4$ &   $9.81$--$9.85$\\
           &                   & semi-tracked  &    $2.20$--$11.38$ &     $0.303$--$0.304$ &      $2.7$--$13.8$ &   $9.78$--$9.81$\\
           &                   & untracked     &    $7.85$--$28.66$ &     $0.351$--$0.352$ &       $1.2$--$4.5$ &   $8.98$--$9.00$\\
\midrule
L-GSC      & charged           & tracked       &     $2.13$--$3.32$ &              $0.219$ &      $6.6$--$10.3$ &           $4.91$\\
           &                   & semi-tracked  &     $1.13$--$2.88$ &              $0.154$ &      $5.3$--$13.6$ &           $4.91$\\
           &                   & untracked     &    $6.10$--$13.50$ &              $0.193$ &       $1.4$--$3.2$ &           $4.91$\\
\midrule
LGSCV      & charged           & tracked       &     $2.40$--$3.74$ &              $0.219$ &       $5.8$--$9.1$ &           $4.91$\\
           &                   & semi-tracked  &     $1.22$--$2.06$ &              $0.154$ &      $7.5$--$12.6$ &           $4.91$\\
           &                   & untracked     &     $2.52$--$4.77$ &              $0.193$ &       $4.0$--$7.6$ &           $4.91$\\
\midrule
GSCodec-S  & charged           & tracked       &     $2.49$--$3.70$ &              $0.219$ &       $5.9$--$8.8$ &           $4.91$\\
           &                   & semi-tracked  &     $1.80$--$3.26$ &              $0.154$ &       $4.7$--$8.5$ &           $4.91$\\
           &                   & untracked     &     $2.93$--$8.04$ &              $0.193$ &       $2.4$--$6.6$ &           $4.91$\\
\midrule
SPZ        & not needed        & tracked       &     $2.80$--$5.21$ &                    --- &              $0.0$ &             ---\\
           &                   & semi-tracked  &     $1.71$--$1.91$ &                    --- &              $0.0$ &             ---\\
           &                   & untracked     &    $5.49$--$13.90$ &                    --- &              $0.0$ &             ---\\
\bottomrule
\end{tabular}
\end{table*}

The ablation of Sec.~\ref{sec:exp5} subtracts this cost from every point
of the concatenated curve of \emph{Flame Steak} and recomputes BD-rate
against the unchanged per-frame curve. BD-rate changes from $-0.4\%$ to
$-4.7\%$ on untracked input, from $-71.1\%$ to $-76.4\%$ on semi-tracked
input and from $-63.4\%$ to $-69.9\%$ on tracked input; the six-sequence average of untracked G-PCC in
Fig.~\ref{fig:avg-grid} is $+0.2\%$.

% ---------------------------------------------------------------------
\section{Concatenated Key Frame Coding}
\label{app:keyframe}

\subsection{D-FCGS}
All D-FCGS results are measured by running its released code on the
3DGStream base stream; no published numbers are used. The first frame of
every GoF is an I-frame coded with FCGS~\cite{chen2025fcgs} at one of five
checkpoints, $\lambda\in\{1,2,4,8,16\}\times10^{-4}$, and every other frame
is predicted from the previous reconstruction by the released motion
network and checkpoint. FCGS reorders the Gaussians of an I-frame and
stores no index map, which breaks the Gaussian correspondence that the
P-frame predictor relies on; run as released, the pipeline reaches only
$14.8$\,dB. We therefore compute, at the encoder, the permutation between
each decoded I-frame and its source and apply it to the P-frames of the
GoF before they are coded. The decoder does not need this permutation,
because the P-frames are reconstructed in the order of the decoded
I-frame. With this alignment, the ground-truth PSNR averaged over the six
sequences is $31.914$\,dB, against $31.91$\,dB published for D-FCGS. On
\emph{Flame Steak} at GoF-$10$, an FCGS I-frame costs $4.86$--$7.69$\,MB, and
I-frames make up $52$--$63\%$ of the total rate of $0.93$--$1.21$\,MB per
frame (Fig.~\ref{fig:app-keyframe}).

\subsection{D-GPCC}
D-GPCC codes all I-frames of a sequence as one concatenated set with G-PCC
($30$ I-frames at GoF-$10$ and $60$ at GoF-$5$), at the five G-PCC rate
points, and keeps the P-frame network, checkpoint and parameters of
D-FCGS. Two properties of G-PCC are compensated at the encoder before the
P-frames are coded. Its output order is aligned as for FCGS. Moreover,
G-PCC does not code the real part of a quaternion; it rewrites each
Gaussian into an equivalent rotation and scale whose largest quaternion
component is the real part. Rendering is unchanged, but the decoded I-frame
would differ from the next source frame by a median rotation of
$1.57$\,rad instead of the true motion of $0.017$\,rad, so the same rewrite
is applied to the P-frames of the GoF. The P-frames of each method are
coded against its own decoded I-frames, and their bitstreams are part of
the rate (App.~\ref{app:note-pframes}).

The overall BD-rate of $-46.2\%$ (Sec.~\ref{sec:exp1}) compares two
six-point curves. GoF-$10$ reaches lower rates and GoF-$5$ higher quality,
so the curve of each method takes its three cheapest points at GoF-$10$ and
its three best points at GoF-$5$: for D-GPCC, the three lowest G-PCC rate
points at GoF-$10$ and, at GoF-$5$, the two highest lossy points and the
point with attributes at unit step; for D-FCGS, $\lambda=16$, $8$ and
$4\times10^{-4}$ at GoF-$10$ and $\lambda=4$, $2$ and $1\times10^{-4}$ at
GoF-$5$. Every point averages the six sequences, with PSNR against the
input averaged over frames.

\subsection{Group lengths of the dynamic codecs}
\begin{figure}[!ht]
\centering
\includegraphics[width=\columnwidth]{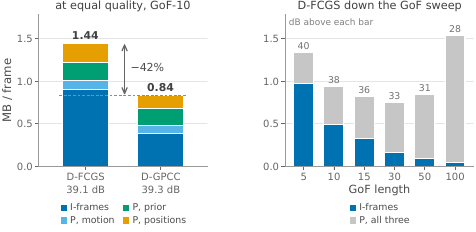}
\caption{What replacing the intra codec moves. Left: the two codecs at
matched quality, six-sequence mean at GoF-$10$ (D-FCGS at
$\lambda=2\times10^{-4}$, D-GPCC at its third G-PCC rate point). The three
P streams move little ($-17\%$; same network and checkpoint, each coded
against its own decoded I-frames), and $84\%$ of the $-42\%$ is the I
segment alone, which falls by $57\%$. Right: the same split down the sweep of
Table~\ref{tab:app-dfcgs-gof}; a longer GoF buys fewer I-frames, and past
GoF-$30$ the P side grows faster than the I side shrinks. Figures above the
bars are MB per frame on the left and PSNR in dB on the right.}
\label{fig:app-keyframe}
\end{figure}

Table~\ref{tab:app-dfcgs-gof} varies the GoF length of D-FCGS on
\emph{Flame Steak}. A longer GoF pays for fewer I-frames, but the motion
network then predicts across more frames and quality drops, so each length
is an operating regime of its own (Fig.~\ref{fig:app-keyframe}, right).
The lowest rate occurs at GoF-$30$, at $33.0$\,dB. The six-sequence runs of D-FCGS and D-GPCC use GoF-$10$, the
released setting, and GoF-$5$, which reaches higher quality.

\begin{table}[!ht]
\centering
\caption{GoF length of D-FCGS on 3DGStream \emph{Flame Steak}, at its
cheapest checkpoint ($\lambda=16\times10^{-4}$). The rate includes all I-
and P-frames; PSNR is measured against the input and averaged over
frames.}
\label{tab:app-dfcgs-gof}
\footnotesize
\setlength{\tabcolsep}{5pt}
\begin{tabular}{@{}rrrr@{}}
\toprule
GoF & I-frames & Rate (MB/f)$\downarrow$ & PSNR (dB)$\uparrow$\\
\midrule
$5$   & $60$ & $1.339$ & $39.86$\\
$10$  & $30$ & $0.935$ & $37.81$\\
$15$  & $20$ & $0.821$ & $36.21$\\
$30$  & $10$ & $0.755$ & $33.02$\\
$50$  & $6$  & $0.839$ & $30.58$\\
$100$ & $3$  & $1.528$ & $27.52$\\
\bottomrule
\end{tabular}
\end{table}

GSCodec-D codes a GoP of frames as HEVC video with inter prediction.
Table~\ref{tab:app-gscodec-gop} varies the GoP length on \emph{Flame
Steak}, with GoP-$1$ as the anchor, just as GoF-$1$ is the anchor in
Table~\ref{tab:gof-full}. Temporal prediction yields almost the entire gain
by GoP-$16$. Beyond that, a longer GoP improves prediction but also widens
the per-attribute value range over which the codec quantizes, and GoP-$30$
gives the best BD-rate. We therefore report GSCodec-D at GoP-$30$; its
released setting is GoP-$16$. GSCodec-D and GSCodec-S differ in more than
grouping: GSCodec-D sorts by Morton code and uses an inter HEVC
configuration, GSCodec-S sorts with PLAS and uses an intra configuration,
and the two configurations differ in eight coding tools and an intra QP
offset. At one frame per group, the static path is $6\%$ cheaper but
$3.0$\,dB worse, a BD-rate of $+11.4\%$, so the distance between the two
codecs in Fig.~\ref{fig:concat-all} does not come from temporal prediction
alone.

\begin{table}[!ht]
\centering
\caption{GoP length of GSCodec-D on 3DGStream \emph{Flame Steak}. Enc\,T is
the wall-clock time of eight parallel HEVC encoders and is not comparable
with Table~\ref{tab:gof-full}; delay is structural, at $30$ frames per
second.}
\label{tab:app-gscodec-gop}
\footnotesize
\setlength{\tabcolsep}{4pt}
\begin{tabular}{@{}ccccc@{}}
\toprule
GoP & \shortstack{BD-rate\,\%\\$\downarrow$} &
\shortstack{Enc\,T\\(s/f)$\downarrow$} & \shortstack{Dec\,T\\(s/f)$\downarrow$} &
\shortstack{Delay\\(s)$\downarrow$}\\
\midrule
$1$   & Baseline  & $8.4$  & $0.7$ & $0.03$\\
$16$  & $-82.80$ & $16.2$ & $0.3$ & $0.53$\\
$30$  & $-85.35$ & $17.2$ & $0.3$ & $1.0$\\
$50$  & $-84.74$ & $12.2$ & $0.3$ & $1.67$\\
$100$ & $-84.82$ & $14.9$ & $0.3$ & $3.33$\\
\bottomrule
\end{tabular}
\end{table}

% ---------------------------------------------------------------------
\section{End-to-End Comparison with Published Results}
\label{app:published}
\begin{figure}[!ht]
\centering
\includegraphics[width=\columnwidth]{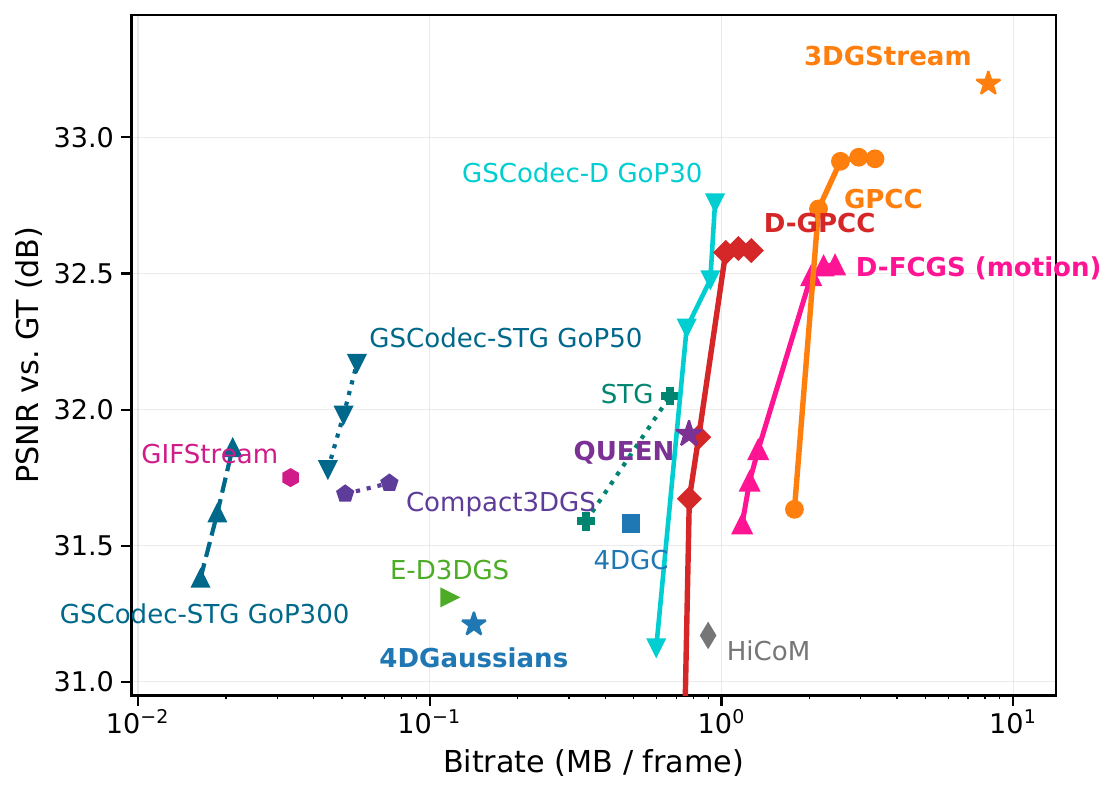}
\caption{Our end-to-end results against the N3DV numbers printed in the
cited papers, all six sequences averaged, quality against the held-out
camera. GPCC and the I-frames of D-GPCC are both the concatenation branch.
D-GPCC and D-FCGS are drawn as the one operating curve the main text uses,
three points at GoF-$10$ and three at GoF-$5$. The stars are estimators at
their own operating point, measured on the representation they train, before
the per-frame export every codec here is given; exporting it costs further
quality. Each point is a complete system, estimator and codec together.}
\label{fig:app-published}
\end{figure}

Every BD-rate in this paper is measured against the input Gaussians, which
isolates compression from estimation (App.~\ref{app:distortion}). That
choice makes our numbers incomparable with the literature, where a method
is scored end to end against the captured views. Fig.~\ref{fig:app-published}
places our runs on that axis instead, next to every N3DV number we could
take from a published table.

Three things limit what the figure can show. First, each row is a complete
system, an estimator and a codec together, so a point that sits low may do
so because of its estimator; our own curves inherit the ceiling of
3DGStream, which Fig.~\ref{fig:app-reference} shows at $33.7$\,dB on one
sequence. Second, the printed values are used exactly as printed: we take
them from the table of the paper that reports them and convert nothing.
Seven of the fourteen methods report a size for the whole $300$-frame
sequence rather than a rate per frame; dividing one by the other would
assume how a shared model is amortized, so those methods appear on the
right of the figure at their quality alone. Third, the quality axis is
narrow: all fourteen published averages lie between $31.15$ and
$32.19$\,dB.

Within those limits, concatenated G-PCC on 3DGStream reaches
$32.9$\,dB, above every published average, but only at $2.5$--$3.4$\,MB per
frame, where the cheapest published methods report under $0.7$. D-GPCC is
the configuration of ours that competes on rate: at GoF-$5$ it reaches
$32.6$\,dB at $1.03$\,MB per frame, above the best published average we
found, QUEEN-l at $32.19$\,dB, at about $1.4$ times its printed size.

One disagreement is visible in the figure and we have not resolved it. Our
run of D-FCGS reproduces its published quality, $31.914$ against
$31.91$\,dB (App.~\ref{app:keyframe}), but not its published rate: we
measure $1.18$--$1.55$\,MB per frame at GoF-$10$ against the $0.46$ printed.
The gap is larger than the I-frame accounting can explain, since the
P-frames alone cost $0.49$--$0.50$\,MB per frame in our run of D-FCGS
(App.~\ref{app:note-pframes}). We therefore compare D-GPCC against our own
run of D-FCGS throughout, under one set of measurement conventions, and
report the effect of concatenation alone by holding the P-frame and
I-frame codecs fixed (App.~\ref{app:note-pframes}); the printed D-FCGS
point is shown here for reference, not as the anchor of any BD-rate.

% ---------------------------------------------------------------------
\section{Per-Sequence Results}
\label{app:perseq}

\subsection{Two rate points that are one}
\label{app:lgsc}
\begin{figure}[!ht]
\centering
\includegraphics[width=\columnwidth]{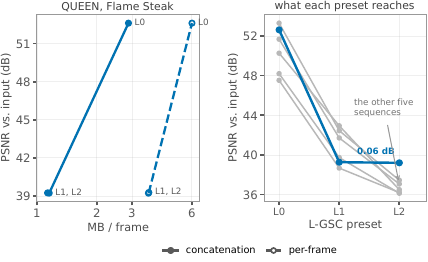}
\caption{Left: on QUEEN \emph{Flame Steak} two of L-GSC's three rate points
land on each other in both branches. Right: what each preset reaches on each
of the six sequences; the last preset costs $2.5$--$6.4$\,dB everywhere
else and $0.06$\,dB here.}
\label{fig:app-lgsc}
\end{figure}

L-GSC has three presets, and its last one changes nothing but the precision
of the higher SH bands, from $6$ to $5$ bits; position, scale, rotation,
opacity and the SH DC term are coded identically at both. On QUEEN input
that step costs almost no rate on any sequence, $0.4$--$1.0\%$, because
QUEEN draws its higher-order SH from a small codebook, $205$--$268$ distinct
vectors for $248$k--$432$k Gaussians in a frame, the same property that makes
the SPZ curves of this input nearly vertical (App.~\ref{app:semitracked}).
On \emph{Flame Steak} it also costs almost no quality: $0.06$\,dB per-frame
and $0.01$\,dB concatenated, against $2.5$--$6.4$\,dB on the other five
sequences (Fig.~\ref{fig:app-lgsc}). Of the six, this sequence has the
narrowest range of higher-order SH values, $3.78$ against $4.23$--$6.14$, and
L-GSC quantizes each attribute over its own range, so one bit less leaves
the smallest error of the six; at $39.2$\,dB that error is below what the
attributes the preset does not touch already contribute.

The two presets are therefore one operating point, in rate and in quality
alike, and a quadratic through three points of which two coincide is fixed
by the gap between them rather than by the data: it dives to $0.014$\,MB per
frame, $80$ times below the cheapest point measured, and returns
$-98.7\%$. We merge rate points that differ by less than $0.25$\,dB before
fitting and use the degree the remaining points support, here a line through
two points, which gives the $-61.2\%$ of Table~\ref{tab:app-perseq}. The
same number follows from the data without any fit: the rate ratios of the
two branches at the three presets are $0.48$, $0.32$ and $0.31$, whose
geometric mean is $-63.9\%$, and the BD-rate with
$\mathrm{PSNR}_{\mathrm{YUV}}$ is $-62.7\%$. Across the repository the rule
changes this one cell of Table~\ref{tab:app-perseq}; the next-smallest gap
between adjacent rate points is $1.05$\,dB, $75$ times wider. The average
column is unaffected, because it averages the $i$-th rate point over the six
sequences first and those means are $3.7$\,dB apart.

\begin{figure}[!ht]
\centering
\includegraphics[width=\columnwidth]{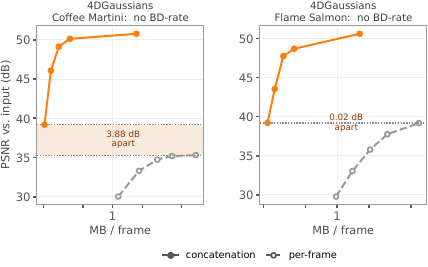}
\caption{The two cells of Table~\ref{tab:app-sweep} that no BD-rate
describes. The two branches never reach the same quality, so there is no
interval to integrate over; the band is the gap.}
\label{fig:app-bdfail}
\end{figure}

Table~\ref{tab:app-perseq} lists the BD-rate of concatenation against
per-frame coding for every codec, correlation pattern and sequence at
GoF-$30$, together with the six-sequence averages of
Fig.~\ref{fig:avg-grid}; as there, the rates of L-GSC, LGSCV and GSCodec-S
include the $4.91$-bit frame-index charge (App.~\ref{app:note-recovery}). On tracked and
semi-tracked input, every codec gains on every sequence. On untracked
input, the sign is the same on all sequences for HGSC and L-GSC (positive)
and for SPZ and GSCodec-S (negative), while G-PCC and LGSCV fall on both
sides of zero. The curves behind every entry are on the project page.

\begin{table*}[t]
\centering
\caption{BD-rate (\%) of GoF-$30$ concatenation against per-frame coding,
per sequence. The last column is the BD-rate between the six-sequence
average curves of Fig.~\ref{fig:avg-grid}, which is not the mean of the
row. CM: \emph{Coffee Martini}, CS: \emph{Cook Spinach}, CRB: \emph{Cut
Roasted Beef}, FSa: \emph{Flame Salmon}, FSt: \emph{Flame Steak}, SS:
\emph{Sear Steak}. $^\dagger$Two of L-GSC's three rate points are one
operating point on this sequence and are merged before the fit, which is
then a line through two points (App.~\ref{app:lgsc}).}
\label{tab:app-perseq}
\footnotesize
\setlength{\tabcolsep}{6pt}
\begin{tabular}{@{}llrrrrrrr@{}}
\toprule
Pattern & Codec & CM & CS & CRB & FSa & FSt & SS & Average\\
\midrule
Tracked      & G-PCC     & $-68.9$ & $-67.6$ & $-64.2$ & $-69.9$ & $-63.4$ & $-62.7$ & $-66.9$\\
(3DGStream)  & HGSC      & $-69.4$ & $-67.4$ & $-62.9$ & $-69.3$ & $-62.5$ & $-62.9$ & $-66.4$\\
             & L-GSC     & $-76.0$ & $-71.1$ & $-68.1$ & $-76.8$ & $-67.0$ & $-68.5$ & $-71.8$\\
             & SPZ       & $-38.4$ & $-50.0$ & $-49.8$ & $-38.2$ & $-48.3$ & $-48.7$ & $-43.0$\\
             & LGSCV     & $-41.8$ & $-44.7$ & $-38.8$ & $-42.6$ & $-39.8$ & $-39.8$ & $-42.0$\\
             & GSCodec-S & $-49.9$ & $-46.5$ & $-46.7$ & $-49.7$ & $-46.2$ & $-46.5$ & $-47.5$\\
\midrule
Semi-tracked & G-PCC     & $-71.4$ & $-69.8$ & $-70.5$ & $-71.8$ & $-71.1$ & $-72.7$ & $-72.2$\\
(QUEEN)      & HGSC      & $-24.1$ & $-19.3$ & $-29.1$ & $-26.6$ & $-34.1$ & $-44.9$ & $-37.9$\\
             & L-GSC     & $-61.9$ & $-64.6$ & $-65.5$ & $-61.3$ & $-61.2^\dagger$ & $-65.9$ & $-63.4$\\
             & SPZ       & $-60.8$ & $-61.0$ & $-61.6$ & $-60.6$ & $-60.8$ & $-61.6$ & $-60.5$\\
             & LGSCV     & $-46.9$ & $-50.4$ & $-49.8$ & $-47.7$ & $-50.0$ & $-51.2$ & $-49.2$\\
             & GSCodec-S & $-43.4$ & $-45.3$ & $-46.6$ & $-43.8$ & $-45.7$ & $-46.8$ & $-45.2$\\
\midrule
Untracked    & G-PCC     & $0.0$   & $+0.9$  & $+1.6$  & $-0.2$  & $-0.4$  & $+0.5$  & $+0.2$\\
(INRIA 3DGS) & HGSC      & $+18.3$ & $+26.1$ & $+23.5$ & $+21.6$ & $+50.4$ & $+28.3$ & $+27.8$\\
             & L-GSC     & $+6.5$ & $+1.1$ & $+0.5$ & $+12.0$ & $+11.7$ & $+0.9$ & $+5.0$\\
             & SPZ       & $-0.1$  & $-0.3$  & $-0.2$  & $-0.1$  & $-0.3$  & $-0.3$  & $-0.2$\\
             & LGSCV     & $+0.8$ & $-2.3$ & $-0.9$ & $-0.1$ & $-9.7$ & $-2.6$ & $-1.4$\\
             & GSCodec-S & $-3.1$ & $-5.1$ & $-4.0$ & $-1.3$ & $-5.4$ & $-4.7$ & $-3.5$\\
\bottomrule
\end{tabular}

\vspace{1.5em}
\caption{BD-rate (\%) of GoF-$150$ G-PCC concatenation against per-frame
G-PCC, per estimator and sequence (abbreviations as in
Table~\ref{tab:app-perseq}). N/A: the PSNR ranges of the two branches do
not overlap. The six-sequence average for 3DGStream is $-69.4\%$. The
\emph{Flame Steak} entries of 4DGaussians, QUEEN and INRIA 3DGS come from
separate runs with identical settings.}
\label{tab:app-sweep}
\footnotesize
\setlength{\tabcolsep}{6pt}
\begin{tabular}{@{}lrrrrrr@{}}
\toprule
Estimator & CM & CS & CRB & FSa & FSt & SS\\
\midrule
3DGStream   & $-71.8$ & $-69.8$ & $-65.9$ & $-73.3$ & $-65.1$ & $-64.9$\\
4DGaussians & N/A     & $-84.0$ & $-88.8$ & N/A     & $-89.5$ & $-92.0$\\
QUEEN       & $-66.3$ & $-61.0$ & $-61.5$ & $-66.0$ & $-62.0$ & $-64.4$\\
INRIA 3DGS  & $-0.9$  & $0.0$   & $+1.1$  & $-1.1$  & $-1.3$  & $+0.5$\\
\bottomrule
\end{tabular}
\end{table*}

Table~\ref{tab:app-sweep} repeats the comparison for G-PCC at GoF-$150$ on
all four estimators, the setting of the right panel of
Fig.~\ref{fig:overview}. The gain again follows the correlation pattern
and is largest for 4DGaussians. Two entries are undefined, and the reason
is a property of the coding unit rather than of the sequence.

G-PCC sizes the position quantizer of a coding unit from that unit's own
bounding box, capping its coded span at $2^{18}$ steps, so a box inflated
by a few far-away points coarsens the grid for everything inside it. The
4DGaussians exporter leaves such points behind (Table~\ref{tab:app-conditioning}), and
what decides how much they cost is whether the unit is split. A single
4DGaussians frame holds $124$k Gaussians, far below the coder's budget of
$1.1$M points per slice, so the per-frame branch codes it as one unit spanning
the whole frame: $6827$ units on \emph{Coffee Martini}, a step of
$2.6\times10^{-2}$. The concatenated GoF-$150$ set holds $18.6$M Gaussians
and is split into $18$ slices, each with its own box; one of them collects
the outliers ($43\,950$ Gaussians, $0.24\%$ of the set, span $6263$) and the
other seventeen have a median span of $440$. \emph{Flame Salmon} splits into
$21$ slices of median span $718$ and at most $1736$, against $4328$ for one
of its frames. Concatenation thus codes the same Gaussians on a grid $6$ to
$15$ times finer, with no change of input scale between the branches.

The per-frame branch therefore saturates at $35.3$ and $39.2$\,dB on these two
scenes even where attributes are coded at quantization step $1$, while the
concatenated branch reaches $50.8$ and $50.6$\,dB (Fig.~\ref{fig:app-bdfail}),
and no BD-rate is defined
between two curves that do not share a PSNR interval. The same effect is
present but mild elsewhere: across the other 4DGaussians scenes the
per-frame ceiling orders with the grid step, $64.8$\,dB at
$3.0\times10^{-4}$, $60.0$ at $1.1\times10^{-3}$, $53.8$ at
$3.5\times10^{-3}$. Leaving the two cells undefined is the conservative
reading: a BD-rate computed there would credit concatenation with repairing
a geometry-quantization artefact rather than with temporal redundancy.

\end{document}